\documentclass[aps,prd,preprint,groupedaddress,showpacs,nofootinbib]{revtex4}

\usepackage{amsmath}
\usepackage{amssymb}
\usepackage{mathtools}
\usepackage{graphicx}
\usepackage{hyperref}
\usepackage{cancel}
\usepackage{enumerate}

\DeclareGraphicsExtensions{.pdf,.png,.jpg,.eps}

\usepackage[applemac]{inputenc}
\usepackage{amsmath}
\usepackage{mathbbol}

\newcommand{\s}{\;\;}

\begin{document}
\unitlength = 1mm

\title{Cosmological singularities and bounce in Cartan-Einstein theory}

\author{Stefano Lucat}
\email{s.lucat@students.uu.nl}

\author{Tomislav~Prokopec}
\email{t.prokopec@uu.nl}

\affiliation{Institute for Theoretical Physics, Spinoza Institute and EMME$\Phi$, Utrecht University,\\
Postbus 80.195, 3508 TD Utrecht, The Netherlands}

\date{\today}

\begin{abstract}
 We consider a generalized Einstein-Cartan theory, in which we add the unique covariant dimension four operators to general relativity that couples 
fermionic spin current to the torsion tensor (with an arbitrary strength). Since torsion is local and non-dynamical, when integrated out it 
yields an effective four-fermion interaction of the gravitational strength. We show how to renormalize the theory, in the one-loop
perturbative expansion in generally curved space-times, obtaining the first order correction to the 2PI effective action in Schwinger-Keldysh ({\it in-in}) formalism. 
We then apply the renormalized theory to study the dynamics of a collapsing universe
that begins in a thermal state and find that -- instead of a big crunch singularity -- the Universe with torsion undergoes a {\it bounce}.
We solve the dynamical equations (a) classically (without particle production); (b) including the production of fermions 
in a fixed background in the Hartree-Fock approximation and (c) including the quantum backreaction of fermions onto the background space-time.
In the first and last cases the Universe undergoes a bounce. The production of fermions due to the coupling to a contracting 
homogeneous background speeds up the bounce, implying that the quantum contributions from fermions is negative, presumably 
because fermion production contributes negatively to the energy-momentum tensor. 
When compared with former works on the subject, our treatment is  fully microscopic
(namely, we treat fermions by solving the corresponding Dirac equations)
and quantum (in the sense that we include fermionic loop contributions). 

\end{abstract}


\maketitle


\section{Introduction}
\label{Introduction}

Elementary particles are described by irreducible representations of the Poincar\'e group. This description is based on the notion of mass, which represents the translational part of the Poincar\'e group, and spin, connected with its rotational part~\cite{Hehl1}. Since general relativity couples gravity to the energy-momentum tensor, which characterise a macroscopic distribution of mass, to the geometrical quantity representing the curvature, it is natural to ask whether also spin could be coupled to the geometry of space-time. The quantity of interest is, in this case, the so-called torsion tensor, which is the antisymmetric part of the connection. The resulting theory was proposed for the first time by E.~Cartan~\cite{Cartan1922,Cartan1923} and in subsequent 
works~\cite{Kibble:1961ba,Sciama:1964wt}, and goes by the name of 
Einstein-Cartan-Kibble-Sciama (ECKS) gravity.

Inclusion of torsion in gravitational theories can be understood as follows: Einstein's general relativity, in first order formalism, has two dynamical variables, the metric $g_{\mu\nu}$ and the connection ${\Gamma^\lambda}_{\mu\nu}$. Postulating that the gravitational action is proportional to the Ricci scalar, leads to the equations of motion, 10 for the metric and 64 for the connection. Assuming that the connection is symmetric, and neglecting the 24 torsion components, leads to the equations of motion for the connection, whose solutions are the Christoffel symbols. However, without this assumption, Einstein's theory will contain torsion which must then be treated as an independent variable. Cartan went as far as arguing that an observer with twisted perceptions can measure a non-vanishing torsion~\cite{Hehl4,Cartan1922}, which implies that there are coordinate transformations leading to non-vanishing torsion. 

Acknowledging these arguments, leads to considering torsion as a dynamical variable. One must therefore identify torsion sources and construct the corresponding torsion-matter interaction terms. This can be done in two ways: by including translational symmetry in the theory, next to diffeomorphisms invariance, or by considering only the coordinate transformation symmetry. 
In this paper we consider the torsion's source to be fermionic matter, which yields to two possibile interactions operators of energy dimension 4\footnote{Here ${S^\lambda}_{\mu\nu} = {\Gamma^\lambda}_{[\mu\nu]}$}
\begin{equation}
\label{eq0}
\sqrt{-g}S_{\lambda\mu\nu} \epsilon^{\lambda\mu\nu\sigma}\left (\xi \bar{\psi}\gamma^5\gamma_\sigma\psi + \xi' \bar{\psi}\gamma_\sigma\psi\right ),
\end{equation}
that respect Lorentz invariance. The particular choice $\xi = 1, \,\xi'=0$ corresponds to the Einstein-Cartan gravity, and can be deduced from the above by imposing translational invariance~\cite{Hehl4}. However we do not know whether translational symmetry is a fundamental component of nature, and we therefore choose to study the most general case (\ref{eq0}). Furthermore an interaction term such as (\ref{eq0}) might follow from the UV completion of gravity\footnote{In Quantum Loop Gravity, for instance, such an interaction does arise with $\xi' = 0$, but $\xi\neq 1$.} and is of interest because it might lead to a classical theory of gravity devoid of singularities. We also would like to point out that in neither of the interaction terms (\ref{eq0}) torsion couples to the fermions spin, since spin, according to the classification in~\cite{PeskinSchroeder}, is given by the spatial part of the tensor matrices of the Clifford basis $\Sigma_i = \frac{i}{4}\epsilon_{ijk}\gamma^{[j}\gamma^{k]} = -\frac{i}{2} \gamma^5\gamma^i\gamma^0$. In our theory torsion couples to the vector and pseudo-vector fermions bilinears.

Since torsion couples to vector currents, its contribution vanishes when averaged on a spatially isotropic distribution of matter, 
and therefore the effects of its interactions are important on small scales. For this reason, ECKS gravity is essentially 
indistinguishable from general relativity on all scales where the latter has been tested, and significantly differs from it only 
at high energies and at small scales. Prominent examples where one can probe high energies and/or small scales are 
cosmology and black holes, the former being the subject of the current study. One should keep in mind 
that, because Cartan theory reduces to general relativity at large scales, 
no experiment so far has been able to disprove Cartan theory~\cite{Bluhm,Kostelecky,Lehnert,Mao:2006bb}, 
which therefore remains a viable microscopic theory of gravity. It is unlikely that torsion can change 
the divergence structure of gravity, thereby gravity with torsion remains non-renormalizable and the question of 
the ultraviolet completion of gravity remains open. This is because torsion in the Cartan-Einstein theory is not a dynamical field, but 
it appears as a Lagrange multiplier and hence it represents a physocal constraint that need not to be separately canonically quantized.

The literature contains several efforts of making predictions using Einstein-Cartan theory, however what all those references have in common is their use of classical spin fluid as a source of torsion~\cite{Poplawski2,Poplawski1,Trautman}. 

Classical spin fluid of Weyssenhoof and Raabe induces a canonical spin tensor density, 
$s_{\mu\nu}^\rho = s_{\mu\nu} u^\rho$, where $s_{\mu\nu}$ 
is the spin density and $s_{\mu\nu}u^\nu = 0$. 
Hehl et al~\cite{Hehl1} point out that "unfortunately,
there seems to be no satisfactory Lagrangian for this
distribution, and therefore no unambiguous road to a
minimally coupled theory".

This classical description is not satisfactory from a field theoretical perspective: in~\cite{Brechet:2007cj,Brechet:2008zz} the torsion tensor in the fluid rest frame is given by ${S^\lambda}_{\mu\nu} = 8\pi G_N s_{\mu\nu} u^\lambda = 2\pi G_N \delta^\lambda_0 \epsilon_{ijk} \bar{\psi} \gamma^5\gamma^k\psi$. This form of the torsion tensor neglects the zeroth component of the axial current which is, as it will become clearer later, the most important contribution to the stress-energy tensor, thus rendering the previous analyses, which are based on
the Weyssenhoof spin fluid, unreliable. 

The main purpose of this paper is to extend the existing classical analysis starting from a microscopic theory, in which no assumptions on the spin fluid are made. We consider the interaction terms (\ref{eq0}) setting $\xi'=0$, which is effectively a generalization of ECKS theory. We then apply this theory to an homogeneous and isotropic universe, initially in a thermal state, undergoing a gravitational collapse and we show that the contribution induced by torsion coupling prevent the formation of singularities. 
Instead, the collapsing universe undergoes a bounce. This conclusion holds both when fermions are treated classically 
--{\it i.e.} when fermion production due to coupling to gravity is neglected -- and quantum field theoretically,
when particle creation due to the fermion coupling to a contracting gravitational background is accounted for.

 The paper is organized as follows. The introductory section~\ref{Introduction} is followed by 
section~\ref{Derivation of the effective theory action and equations of motion},
in which we show how to construct a theory with non-vanishing torsion and how one can integrate out torsion 
to formulate an effective field theory in general. In the case of fermions, this yields to an effective four-fermion interaction, 
involving the square of the pseudo-vector current, $\bar{\psi}\gamma^5\gamma^\mu\psi$~\cite{Kasuya1977}. 
We then perform a Wick contraction to derive a 2-loop effective action and 2-loop effective energy-momentum tensor. 
In section~\ref{2PI effective action renormalization}
we perform dimensional regularization to remove ultraviolet divergences from the energy-momentum tensor
(the details of the procedure are rendered to appendices~\ref{Propagators.Expansion} and~\ref{Mode Functions}). 
In section\ref{Analytical solution in the massless regime} we solve the Einstein's equations for the scale factor,
{\it i.e.} the Friedmann equations, and the fermionic field equations self-consistently
(assuming light fermions).
In section~\ref{Self-consistent backreaction} the semiclassical equations for gravity plus matter 
are numerically solved and the results are plotted taking for the background fluid radiation and dust.
Our result show that quantum effects increase the torsion backreaction, such that the bounce occurs 
earlier than in the classical case.
 section~\ref{Validity of the semiclassical treatment} is devoted to a discussion of the range of validity of
our semiclassical approximation.
We conclude in section~\ref{Conclusion and discussion} with final remarks and comments.


\section{Derivation of the effective theory action and equations of motion}
\label{Derivation of the effective theory action and equations of motion}

As in general relativity, the Einstein-Cartan action is the Einstein-Hilbert action, but 
expressed in terms of the general affine connection containing torsion~\footnote{We work in natural units, $\hbar = c = 1$.},
\begin{equation}
\label{eq1.1}
S = \int \text{d}^D x\sqrt{-g} \left ( \mathcal{L}_\textit{m}(g,\Gamma) - \frac{1}{16 \pi G_N} R(g,\Gamma) \right )\,,
\end{equation}
such that the dynamical variables of the theory are the metric $g_{\mu\nu}$ (or more precisely the tetrad $e^a_\mu$) 
and the torsion tensor ${S^\lambda}_{\mu\nu} = \Gamma^\lambda_{[\mu\nu]}$. 
In~(\ref{eq1.1})  $R$ denotes the Ricci scalar, $G_N$ is the Newton constant and $d^Dx$ is the infinitesimal
$D$-dimensional volume element (we keep $D$ general for the purpose of dimensional
regularization utilized in this work).
Varying the action (\ref{eq1.1})
 with respect to the tetrad field yields to the modified Einstein's equations~\cite{Arkuszewski1974,Hehl1},
\begin{equation}
\label{eq1.2}
\begin{split}
G_{\mu\nu} =\s & 8\pi G_N \,T_{\mu\nu} - \overset{\circ}{\nabla}_\kappa (2{S_{(\mu\nu)}}^\kappa+{S^\kappa}_{\mu\nu}) + 2\overset{\circ}{\nabla}_\nu S_\mu - 2  \overset{\circ}{\nabla}_\kappa S^\kappa g_{\mu\nu} \\
-& 2 S_\kappa (2{S_{(\mu\nu)}}^\kappa+{S^\kappa}_{\mu\nu}) + (2{S_{(\kappa\nu)}}^\lambda + {S^\lambda}_{\kappa\nu} )(2{S_{(\mu\lambda)}}^\kappa+{S^\kappa}_{\mu\lambda}) \\
-& \frac{1}{2} g_{\mu\nu} \left [4 S_\kappa S^\kappa + (2 S_{(\kappa\rho)\lambda}+S_{\lambda\kappa\rho})(2 S^{(\rho\lambda)\kappa}+S^{\kappa\rho\lambda})\right ] 
\,,
\end{split}
\end{equation}
while the variation of the action~(\ref{eq1.1}) with respect to the torsion leads to the Cartan equation
\begin{equation}
\label{eq1.3}
     4 g_{\lambda[\mu}S_{\nu]} + S_{\lambda[\mu\nu]}- 2 S_{[\mu\nu]\lambda} =  8 \pi G_N\, \Pi_{\lambda[\mu\nu]}
\, ,
\end{equation}
where 
\begin{equation}
\label{eq1.4}
\Pi_{\lambda[\mu\nu]} = \frac{2}{\sqrt{-g}}\frac{\delta \mathcal{L}_\textit{m} }{\delta S^{\lambda\mu\nu}}
\,
\end{equation}
and $S_\mu={S^\alpha}_{\mu\alpha}=-{S^\alpha}_{\alpha\mu}$ is the trace of the torsion tensor.
The anti-symmetric terms on the right hand side of Eq.~(\ref{eq1.2}) vanish due to the covariant conservation 
(with respect to the full metric) of the Noether current associated with rotations~\cite{Hehl1}, 
$\nabla_\lambda\tau^\lambda_{\;\mu\nu}=0$, where 
${\tau^\lambda}_{\mu\nu}={S^\lambda}_{\mu\nu}-2 \delta^\lambda_{[\mu}S_{\nu]}$.

The Cartan equation~(\ref{eq1.3}) is a purely algebraic relation, which can be solved to derive an
 effective action for the theory. Note that this also implies that the torsion field does not propagate, 
but can only exist inside a matter distribution. The general effective action is 
\begin{eqnarray}
\label{eq1.5}
S_\textit{\rm eff} &=& \int \text{d}^D x \sqrt{-g}\overset{\circ}{\mathcal{L}}_\textit{m}
 - \int \text{d}^D x \frac{1}{16 \pi G_N} \sqrt{-g} 
\nonumber\\
&\times & \bigg ( \overset{\circ}{R} + 8 \pi G_N\xi\left (\overset{\circ}{\nabla}_\kappa  - 4 \pi G_N \xi \Pi_\kappa\right ) \Pi^\kappa
 - \left (8 \pi G_N\xi\right)^2 \Pi_{\rho\mu\kappa} \Pi^{\mu\kappa\rho} \bigg ) 
\,,
\end{eqnarray}
where the superscript $\circ$ indicates quantities defined by the usual Christoffel symbols from general relativity
and $\Pi_\kappa={\Pi^\alpha}_{\kappa\alpha}$ denotes the trace of the torsion source $\Pi_{\alpha\mu\nu}$.

One question that might arise at this stage is, what sort of fields couple to torsion and its source $\Pi_{\lambda\mu\nu}$? 
Scalar fields do not couple to torsion {\it via} the kinetic term, however 
a non-minimal coupling to gravity, such as $\xi \phi^2 R$, can source torsion.
Next, gauge fields can also couple to torsion, but the prize to pay is 
a breakdown of gauge invariance.~\footnote{Indeed, the field strength in spaces with torsion is,
$F_{\mu\nu}=\nabla_\mu A_\nu - \nabla_\nu A_\mu =   \overset{\circ}F_{\mu\nu}-2{S^\alpha}_{\mu\nu}A_\alpha$. 
The last term will induce in the gauge field action a quadratic contribution in $A_\mu$ that does not respect gauge symmetry. 
Thus, varying the gauge field action with respect to the torsion tensor will generate a non-vanishing torsion source tensor.
It would be of interest to investigate physical effects of such a torsion source.
} 
In this paper we study the case in which fermions couple to torsion, which are abundantly present in the standard model,
and  whose interactions with gravity are modified according to ECKS theory,
which is obtained from the more general effective theory~(\ref{eq1.5}) by inserting the suitable fermionic source,
which is what we discuss next. 

We consider $\mathcal{L}_\textit{m}$ to be the Dirac Lagrangian in curved space-time~\cite{Hehl1},
\begin{equation}
\label{eq1.6}
\begin{split}
\sqrt{-g}\mathcal{L}_\psi =&\sqrt{-g} \bigg [ \frac{i}{2} \left ( \bar{\psi} \gamma^\mu \psi_{,\mu} - \bar{\psi}_{,\mu}  \gamma^\mu \psi + \bar{\psi} \{\gamma^\mu,\overset{\circ}{ \Gamma}_\mu\} \psi \right ) \\
&\hskip 1cm 
- m_R \bar{\psi} \psi - i m_I \bar{\psi} \gamma^5\psi
 - \frac{1}{4} S_{\lambda\mu\nu}\epsilon^{\lambda\mu\nu\sigma} \left (\xi \bar{\psi}\gamma^5\gamma_\sigma\psi
 + \xi' \bar{\psi}\gamma_\sigma\psi \right) \bigg ],
\end{split}
\end{equation}
where we introduced an extra CP violating interaction, proportional to $m_I$~\cite{Prokopec4}. These type of interaction can generate CP-violating effects only if  $m_I$ is space or time dependent
(which can be {\it e.g.} obtained if $m_I$ is generated by a scalar field condensate), or when the background is space 
and/or time dependent. If that is not the case, then the $m_I$ term can be simply removed by 
a $\gamma^5$ dependent global rotation of the fermionic field.

The interactions terms proportional to $\xi$ and $\xi'$ will yield to similar effects, although the pseudo-vector current 
might contain additional CP violations. Therefore we are going to set $\xi' = 0$, which leads to a simpler theory. 
The complementary case -- in which $\xi^\prime\neq 0$, $\xi=0$ --
is discussed in appendix~\ref{appendix A}. 

The spin density coupling to torsion is, 
\begin{equation}
\label{eq1.7}
 \Pi^{\mu\nu\lambda} = - \sqrt{-g} \frac{i \xi}{4} \bar{\psi} \gamma^{[\mu} \gamma^\nu \gamma^{\lambda]} \psi = - \frac{ \xi}{4} \epsilon^{\mu\nu\lambda\sigma} \bar{\psi} \gamma^5 \gamma_\sigma \psi.
\end{equation}
This equation means  that the torsion source $\Pi^{\mu\nu\lambda}$ is totally antisymmetric, which then implies that 
the torsion tensor in~(\ref{eq1.4}) is also totally antisymmetric (skew-symmetric). This has implications in the motion of test particles: the geodesic equation, very well tested in the solar system and light bending experiments, was formulated using the Christoffel connection. Introducing torsion leads to an ambiguity in the choice of geodesic equation: namely, particles in space-times with torsion would neither follow path of minimal length nor auto parallel path, but trajectories determined by their own equations of motion. Instead, the torsion contribution forces particles to follow trajectories in between these two definitions. This ambiguity, however, is not present if the torsion tensor 
is completely antisymmetric, in which case the two different equations just reduce to the same equation~\cite{Hehl1}. 

Having solved the Cartan equation, we can plug its solution back to the effective action~(\ref{eq1.5})
and derive the effective theory known as Einstein-Cartan-Kibble-Sciama (ECKS) gravity. This gives~\cite{Kasuya1977},
\begin{equation}
\label{eq1.8} 
S = \int \text{d}^D x \sqrt{-g} \left (\overset{\circ}{\mathcal{L}}_\psi - \frac{1}{16 \pi G_N} \overset{\circ}{R} +  \frac{3 \pi G_N  \xi^2}{2} \bar{\psi} \gamma^5 \gamma^\sigma \psi \bar{\psi} \gamma^5 \gamma_\sigma \psi\right ).
\end{equation}
Note that torsion has completely disappeared from ECKS gravity
and all of its contributions are now encoded in the effective four-fermion interaction as given in~(\ref{eq1.8}). 
Since the dimensionless coupling $\xi$ is arbitrary, the strength of this new effective four-fermion interaction 
is unspecified (a rough estimate of the available experimental constraints restrain $\xi\leq 10^{30}$, see~Ref.~\cite{Lehnert}).

The energy momentum tensor implied by ECKS theory~(\ref{eq1.8}) is~\cite{Poplawski3},
\begin{equation}
\label{eq1.9} 
\begin{split}
T_{\mu\nu} = &\frac{i }{2} \:(\bar{\psi} \gamma_{(\mu} D_{\nu)} \psi - D_{(\nu} \bar{\psi} \gamma_{\mu)}  \psi ) 
+ g_{\mu\nu}   \frac{3 \pi G_N \xi^2}{2} \bar{\psi} \gamma^5 \gamma^\sigma \psi \bar{\psi} \gamma^5 \gamma_\sigma \psi
\end{split}
\end{equation}
and the Dirac equation reads
\begin{equation}
\label{eq1.10} 
\left (i \gamma^\mu D_\mu   - m_R - i m_I \gamma^5 \right)\psi 
   = -(3 \pi G_N  \xi^2) (\bar{\psi} \gamma^5 \gamma^\sigma \psi) \gamma^5 \gamma_\sigma \psi
\,,
\end{equation}
where the term in parentheses on the right-hand-side is a scalar (all spinorial indices in parentheses are summed over).
Together with the effective Dirac equation~(\ref{eq1.10}) the semiclassical Einstein equations,
\begin{equation}
 G_{\mu\nu} =8\pi G_N\langle T_{\mu\nu}(x) \rangle 
\,,
\label{semiclassical Einstein}
\end{equation}
constitute the set of equations we solve in this paper, where 
$\langle T_{\mu\nu}(x) \rangle \equiv {\rm Tr}[\rho_{\rm H}T_{\mu\nu}(x)]$ is the expectation value of 
the stress-energy operator (in Heisenberg picture) taken with respect to some density operator $\rho_{\rm H}$ 
(also in Heisenberg picture).
But before we proceed, we need to make sure that all the quantities in~(\ref{eq1.9}--\ref{semiclassical Einstein})
are well defined, {\it i.e.} finite, and that is the question we address in the next section. For technical reasons,
it turns out easier to solve the (perturbatively expanded) two-particle irreducible (2PI) equation for two-point 
functions rather than the Dirac equation~(\ref{eq1.10}). That is also the reason why we perform both renormalization 
of the 1PI effective action (for the Dirac field) as well as of the corresponding 2PI effective action.


\section{2PI effective action and its renormalization}
\label{2PI effective action renormalization}

In this section we address the problem of divergences arising from the vacuum expectation value of the energy momentum tensor, and how to deal with them. Before we begin our analysis, however, a comment is in order: the interaction term produced by integrating the torsion field, in Eq. (\ref{eq1.8}), is a four-fermion vertex. It is a well known fact that such theories are, in general, non-renormalizable. This is because the coupling constant of this interaction, {\it i.e.} $3\pi G_N\xi^2/2$, is dimensionful. 
This will then generate infinitely many divergent inequivalent diagrams, that cannot be absorbed by the addition of finitely many counter-term, as can be shown by the power counting argument, meaning that the theory is non renormalizable. However, the theory under consideration here is an effective one. The closest analogue is the Fermi theory of electromagnetism: at energies much lower than the electron mass~($m_e$), one can integrate out the gauge field of the photon, to construct an effective theory with four-fermion interactions. The underlying assumption of this procedure is that high energy physics decouples from low energy, such that the effective theory makes the correct predictions up to the cutoff scale $m_e$. When calculating the vacuum expectation value of operators, however, one encounters divergences. But, since the theory is not UV complete, one can imagine cutting the momentum integrals~(which give the divergences in the limit $k\rightarrow \infty$) at a finite cutoff scale $\Lambda$. As long as $\Lambda$ remains smaller than the cutoff scale of the effective theory, the vacuum expectation value obtained should be trusted, since up to the cutoff scale one trusts the effective theory. 
Scale dependence will appear, however, in the operator vacuum expectation values.

The procedure described above corresponds to cutoff regularization. In this paper, however, 
we will use the more powerful tool of dimensional regularization (that respects all the symmetries of the underlying theory), 
with the following caveat: since the result of the renormalization procedure should not depend on the scheme used, the scale dependence that we will find is the same one would have found in cutoff regularization. It therefore represents the scale dependence of the vacuum expectation values of operators in the effective theory. To improve the results one could use the renormalization group techniques, which describe even more accurately the scale dependence found in the vacuum expectation values of operators, but that is not needed for our purposes. In appendix~\ref{Mode Functions} we give an example of the universality of renormalization we just mentioned, by constructing a perturbative mass expansion and emplying mode functions renormalization. We find the same divergent structure we derive in this section, however, since we do not re-sum the perturbation theory, the results of appendix~\ref{Mode Functions} and this section differ by a constant. 
\begin{figure}
\centering
\includegraphics[width=\textwidth]{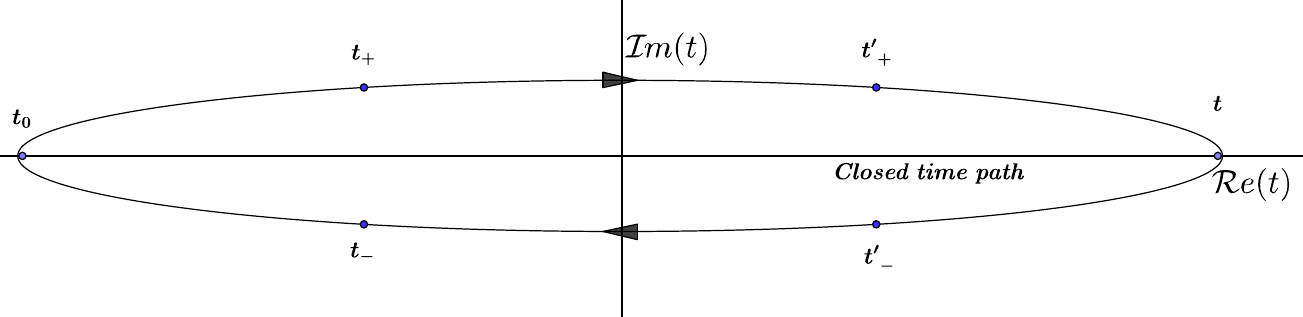}
\caption{The closed time path appearing in the propagator~(\ref{polarities propagator}) in the Schwinger-Kedysh formalism. 
The closed time contour implies that, when propagating from $t$ to $t'$, there are four possible choices. 
The propagators in the Feynman diagrams acquire a direction and, for a given diagram containing non-equal times, 
the propagator has to be summed over all component two-point functions in 
Eqs.~(\ref{eq6.2.1}--\ref{eq6.2.4})~\cite{Keldysh}.}
\label{fig:Image2PIbis}
\end{figure}

In solving the Dirac equation~(\ref{eq1.11})  we are going to consider an initial thermal state for the field $\chi=a^{(D-1)/2}\psi$ 
and evolve it according to the Schwinger-Keldysh~\cite{Chou, Keldysh, Schwinger} formalism. 
The Schwinger-Keldysh formalism is generally used to describe the evolution of operators in out-of-equilibrium quantum field theory.
The initial equilibrium state can be characterized by requiring that macroscopic observables do not depend on time~(in absence of the expansion of the universe). 
In the microscopic theory this amounts to requiring detailed balance~\footnote{Detailed balance means that, 
given two microscopic states $A$ and $B$, 
the transition probability $\mathcal{P}$ satisfies $\mathcal{P}(A\rightarrow B) = \mathcal{P}(B\rightarrow A)$.} to be satisfied, 
such that the macroscopic system stays in the same state at all times~\cite{Chou}. 
Of course, in our case of a contracting universe, the initial thermal state will evolve to a non-thermal state that 
violates detailed balance, making the Schwinger-Keldysh formalism the natural choice.

The fundamental objects used to build perturbation theory in the Schwinger-Keldysh formalism
are the following two-point functions~\cite{LeBellac,Prokopec4},
\begin{subequations}
\begin{eqnarray}
\label{eq6.2.1} iS^{++}_{\alpha\beta}(x; \,x') &=& \langle \mathcal{T}\big [ \psi_\alpha (x) \bar{\psi}_\beta (x')\big ]\rangle\,, \\
\label{eq6.2.2} iS^{-+}_{\alpha\beta}(x; \,x') &=& \langle \psi_\alpha (x) \bar{\psi}_\beta (x')\rangle\,, \\
\label{eq6.2.3} iS^{+-}_{\alpha\beta}(x; \,x') &=&- \langle \bar{\psi}_\alpha (x) \psi_\beta (x')\rangle\,, \\
\label{eq6.2.4} iS^{--}_{\alpha\beta}(x; \,x') &=& \langle \bar{\mathcal{T}}\big [\psi_\alpha (x) \bar{\psi}_\beta(x')\big ]\rangle\,, 
\end{eqnarray}
\end{subequations}
where $\mathcal{T}$($\bar{\mathcal{T}}$) denote the (anti-)time ordering operation
and, as before, $\langle\,\cdot\,\rangle \equiv {\rm Tr}[\rho_{\rm H}\,\cdot\,]$ denotes the expectation value 
with respect to the density operator $\rho_{\rm H}$ in Heisenberg picture. 
Eqs.~(\ref{eq6.2.2}--\ref{eq6.2.3}) correspond to the positive and negative frequency Wightman functions, respectively, and 
Eqs.~(\ref{eq6.2.1}) and~(\ref{eq6.2.4}) are the Feynman (time-ordered) and Dyson (anti-time-ordered) propagators,
which can be expressed in terms of the Wightman functions as follows,
\begin{eqnarray}
 iS^{++}_{\alpha\beta}(x; \,x') &=& \theta(t\!-\!t') iS^{-+}_{\alpha\beta}(x; \,x')
                          +\theta(t'\!-\!t) iS^{+-}_{\alpha\beta}(x; \,x')
\,,\nonumber\\
 iS^{--}_{\alpha\beta}(x; \,x') &=& \theta(t\!-\!t') iS^{+-}_{\alpha\beta}(x; \,x')
                          +\theta(t'\!-\!t) iS^{-+}_{\alpha\beta}(x; \,x')
\,,
\label{Feynman and Dyson}
\end{eqnarray}
where $\theta(x)$ is the Heaviside $\theta$-function. 
The 2-point functions in 
Eqs.~(\ref{eq6.2.1}--\ref{eq6.2.4}) can be connected with the propagator in this formalism, namely 
\begin{equation}
 iS_{\alpha\beta}(x; \,x') = \langle \mathcal{T}_\mathcal{C} [ \psi_\alpha (x) \bar{\psi}_\beta (x')]\rangle=\begin{pmatrix} 
iS^{++}_{\alpha\beta}(x;x') & iS^{+-}_{\alpha\beta}(x;x') \\
iS^{-+}_{\alpha\beta}(x;x') & iS^{--}_{\alpha\beta}(x;x')
\end{pmatrix}\,,
\label{polarities propagator}
\end{equation}
%
%
%
where $\mathcal{C}$ is the complex contour shown in Figure \ref{fig:Image2PIbis}. Then points taken on the upper or lower branch of the contour ($t_\pm,\,t'_\pm$) will be ordered according to the standard time or anti-time ordering, while points lying on different branches will be automatically ordered. From this we deduce the ``$-$'' sign in Eq. (\ref{eq6.2.3}), due to the anti-commutation relations for fermions.

On the right hand side of Eq.~(\ref{polarities propagator}) we have introduced the matrix~(Keldysh) notation for the Schwinger-Keldysh propagator, which is what we will use in the following. Note that, when not specified, we write $iS(x; \,x')$ for the full matrix in Eq.~(\ref{polarities propagator}). Sometimes we will need to consider $\sigma_z iS(x; \,x')$, where $\sigma_z={\rm diag}(1,-1)$ 
only acts on the Schwinger-Keldysh matrix structure.
The propagator~(\ref{polarities propagator}), or equivalently its components~(\ref{eq6.2.1}--\ref{eq6.2.4}), can be used 
to construct perturbation theory for approximate calculations of expectation values of operators. In this paper we shall use it to 
calculate the two-loop expectation value of the stress-energy tensor. 
When the propagator~(\ref{polarities propagator}) is used, then  the perturbation theory is formally identical to the standard
perturbation theory. However, in concrete applications it is often more convenient to use its 
components~(\ref{eq6.2.1}--\ref{eq6.2.4}), in which case the perturbation theory looks like the standard perturbation
theory, but in addition each vertex acquires a $+$ or $-$ polarity, and in order to calculate a Feynman diagram, 
one has to sum over all possible polarities that appear on internal legs. We shall now illustrate how this works 
in practice by considering  perturbation theory associated with the ECKS action~(\ref{eq1.8})
 in the Schwinger-Keldysh (or {\it in-in}) formalism. We begin by constructing the corresponding 2PI  effective action.

\begin{figure}
\centering
    \includegraphics[width=5in]{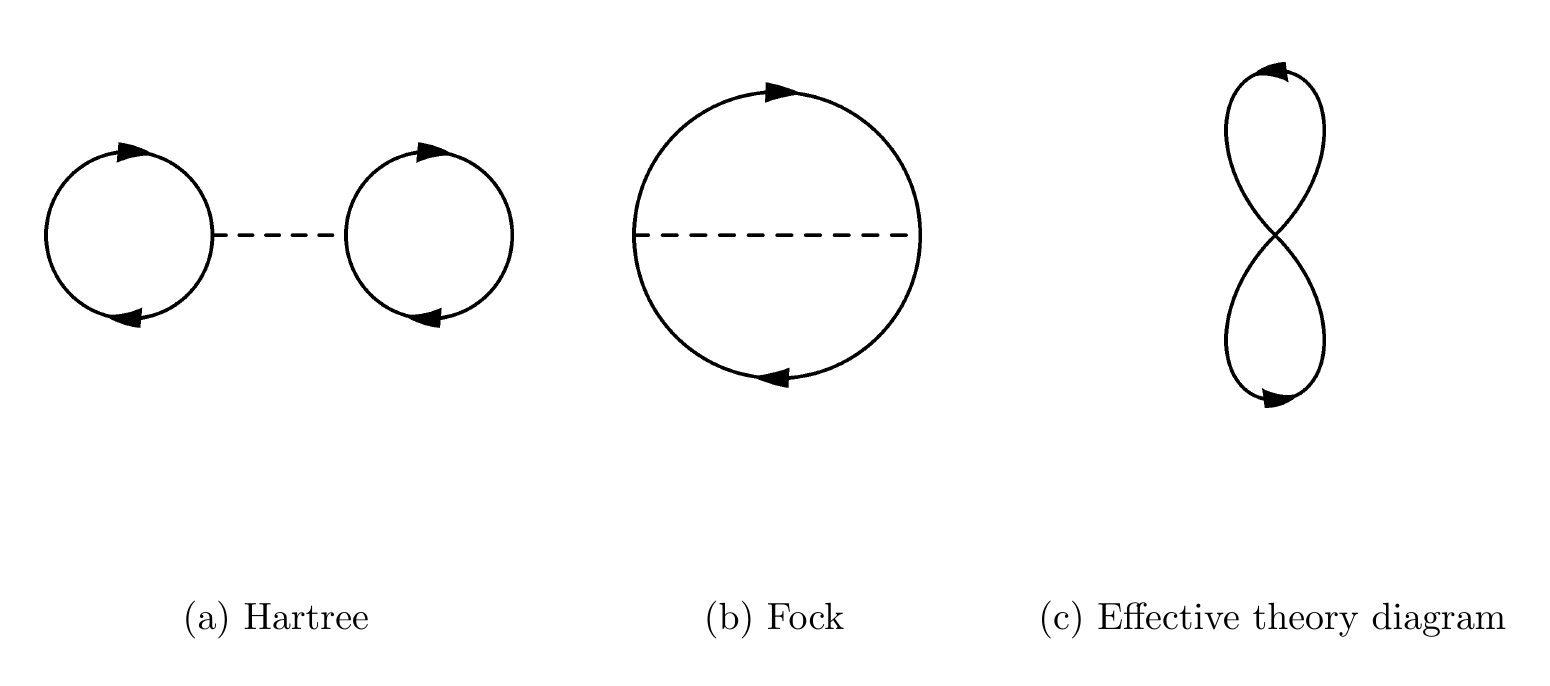}
\caption{The two-loop diagrams that  
graphically represent the Hartree (a) and Fock (b) terms in the effective action~\ref{2PI action: 2 loop}. 
The dashed lines corresponds to the torsion field which reduces to a point in the effective theory. 
The diagram (c) represents both the (a) and (b) diagrams. 
The effective theory diagrams represent the two different Wick contractions of the spin indices 
given in Eq.~(\ref{eq1.15}).}
\label{fig:fig2}
\end{figure}

 In general the 2PI effective action can be written as~\cite{PhysRevD.10.2428,Prokopec4, Calzetta},
\begin{eqnarray}
 \Gamma[S,g_{\mu\nu}] &=&\int\!\text{d}^D x \sqrt{-g}
\!\Big[\text{Tr}[\sigma_z \left (i\cancel{D}\! -\! m_R\! -\! im_I\gamma^5 \right )(i S)]\!-\! \frac{i}{\sqrt{-g}} \text{Tr}\log  (i S)\Big]
\nonumber\\
 &+&
 \Gamma_2[i S,g_{\mu\nu}] \,,\quad
\label{eq2PI.7}
\end{eqnarray}
where the trace is taken on spinorial indices, 
$(iS)$ is the full fermion propagator in the Schwinger-Keldysh formalism
and  $\Gamma_2$ is the sum of all 2PI diagrams containing two or more loops. 
Here we are interested in the two-loop contributions, which consist of 
the Hartree and Fock terms shown in figure~\ref{fig:fig2}, and can be written as, 
\begin{eqnarray}
 &&\Gamma_2[iS,g_{\mu\nu}] =\!\!\nonumber\\
  &=& \!\!\frac{3 \pi G_N \xi^2}{2}\!\int\!\text{d}^D x \sqrt{-g}\sum\limits_{a,b = \pm}\sigma^{ab}_z
   \Big[ {\rm Tr}\big[iS^{aa}(x;x)\gamma^5\gamma^\sigma\big]
       {\rm Tr}\big[iS^{aa}(x;x)\gamma^5\gamma_\sigma\big]
\nonumber\\
       &&\hskip 3.5cm
   - {\rm Tr} \big[iS^{aa}(x;x)\gamma^5\gamma^\sigma iS^{aa}(x;x)\gamma^5\gamma_\sigma\big]\Big]\nonumber \\
   \label{2PI action: 2 loop}&\equiv&  \!\!\frac{3 \pi G_N \xi^2}{2}\!\int\!\text{d}^D x \sqrt{-g}{\rm Tr} \left\{
   \Big[\sigma_z i\Pi(x,x) iS(x;x) \Big]\right\}
\,,
\qquad
\end{eqnarray}
where we defined the self-energy matrix as 
\begin{equation}
\begin{split}
\label{self-energy} i\Pi(x,x) =& \begin{pmatrix}
{\rm Tr} \left (iS^{++}(x,x)\gamma^5\gamma^\sigma\right )\gamma^5\gamma_\sigma& 0 \\
0& {\rm Tr} \left (iS^{--}(x,x)\gamma^5\gamma^\sigma\right )\gamma^5\gamma_\sigma 
\end{pmatrix} \\
&-\begin{pmatrix}
\gamma^5\gamma^\sigma iS^{++}(x,x)\gamma^5\gamma_\sigma & 0 \\
0&\gamma^5\gamma^\sigma iS^{--}(x,x)\gamma^5\gamma_\sigma 
\end{pmatrix}
\end{split}
\end{equation}
This action can be derived by making use of a double Legendre transform~\cite{PhysRevD.10.2428}. 
A simpler derivation can be executed by noting that the form of the action~(\ref{2PI action: 2 loop})
 is that of the standard perturbative two-loop action associated with the ECKS effective action~(\ref{eq1.8}),
which can be obtained by applying the Wick theorem on the four-point interaction in the ECKS action,
\begin{equation}
\label{eq1.15} \langle\bar{\psi} \Gamma\psi \bar{\psi} \tilde{\Gamma} \psi \rangle 
 =\text{Tr}\big (\langle\bar{\psi} \Gamma \psi \rangle\big )\text{Tr}\big (\langle\bar{\psi}\tilde{\Gamma} \psi\rangle\big) 
- \text{Tr}\left(\langle\bar{\psi} \psi\rangle\Gamma\langle\bar{\psi} \psi\rangle\tilde{\Gamma} \right )
\,,
\end{equation}
where $\Gamma\, , \tilde{\Gamma}$ denote any two elements of the Clifford algebra
(in (\ref{2PI action: 2 loop}) $\Gamma=\gamma^5\gamma^\sigma$, $\tilde{\Gamma}=\gamma^5\gamma_\sigma$).
The first term in Eq.~(\ref{eq1.15}) is the Fock term and the second is the Hartree term. 
These contributions are illustrated as the Feynman diagrams in Figure~\ref{fig:fig2}, 
and  they contribute as two-loop diagrams in the effective action and in the energy-momentum tensor
and as one-loop diagrams in the equations of motion.

The detailed procedure to characterise the divergent terms in the effective action~(\ref{eq2PI.7}) is described in appendix~\ref{Propagators.Expansion}. 
Here we just quote the form of the counter-terms action, which is required to remove the divergences appearing in Eq.~(\ref{eq2PI.7}). 
As shown in appendix~\ref{Propagators.Expansion}, we ought to add the following counter-terms, for perturbative renormalization of the effective action,
\begin{eqnarray}
S^{(\text{ct})}[S, g_{\mu\nu}]  \!=\! \int\!\!\text{d}^D x \sqrt{-g} \sigma_z \bigg [
   \frac{\Delta\Lambda}{16\pi G_N}\! -\! \frac{\Delta G_N^{-1}}{16\pi } R\! +\! \Delta\alpha R^2
 \! -\!\Big[\Delta m_R+\Delta\beta R(x)\Big]\big(iS(x;x)\big) \bigg ],\quad
\label{eq2PI.1} 
\end{eqnarray}
Requiring that the relations~(\ref{square completion relations}) between the counter-terms coupling constants,
given in appendix~\ref{Propagators.Expansion}, leads to our main result for this section, the renormalized 2PI effective action, which up to two loops reads~(\ref{Action and counterterms: end}), 
\begin{eqnarray}
\label{eq2PI.11}
 \Gamma_{\textrm{ren}}[i\tilde{S}_0^{ab}, g_{\mu\nu}] 
&=& \int\!\text{d}^D x \sqrt{-g} \sigma_z\Bigg \{
   \frac{\Lambda^{\rm (ren)}}{16\pi G_N}\! -\! \frac{({G_N^{\rm (ren)}})^{-1}}{16\pi } R\! +\! \alpha^{\rm (ren)} R^2 \!+\! \zeta^{\rm (1), (ren)} R_{\mu\nu}R^{\mu\nu}\nonumber \\
     &&  \!+\! \beta^{\rm (ren)} R \big(i{\tilde S}^{\rm(reg)}(x;x)\big) 
      +\text{Tr}[ \left (i\cancel{D}\! -\! m^{\rm (ren)}_R\! -\! im_I\gamma^5 \right ) (i{\tilde S}^{\rm(reg)}(x;x))]
      \nonumber\Bigg\}\\
      &&+\frac{3 \pi G_N \xi^2}{2}\int\!\text{d}^D x \sqrt{-g} {\rm Tr} \left\{
   \sigma_z i\Pi^{\rm(reg)}(x,x) iS^{\rm(reg)}(x;x) \right\}
\,.
\end{eqnarray} 
where the fermion propagator is still a matrix, according to~(\ref{polarities propagator}), and the renormalized coupling constants are given by 
\begin{eqnarray}
\label{renormalized coupling constants}
 \frac{\Lambda^{\rm (ren)}}{16\pi G_N}&=&- \frac{3 \pi G_N \xi^2 |m|^6}{128 \pi^4}\bigg (  \gamma_E + \log \frac{|m|^2}{4\pi\mu^2}\bigg )^2+\frac{\Lambda^{\rm b}}{16\pi G_N}  \,,\qquad\nonumber\\
\alpha^{\rm (ren)} &=& -\frac{ \pi G_N \xi^2|m|^2}{6'144 \pi^4} \bigg (  \gamma_E - 1 + \log \frac{|m|^2}{4\pi\mu^2}\bigg )^2+\alpha^{\rm b} \,,\qquad\nonumber\\
\frac{( {G_N^{\rm (ren)})}^{-1}}{16\pi } &=& \frac{ \pi G_N \xi^2|m|^4}{256\pi^4} \bigg (  \gamma_E - 1 + \log \frac{|m|^2}{4\pi\mu^2}\bigg )\times\bigg (  \gamma_E + \log \frac{|m|^2}{\mu^2}\bigg )  +\frac{( {G_N)}^{-1}}{16\pi } \,,\nonumber\\
\beta^{\rm (ren)}&=& -\frac{ \pi G_N \xi^2|m|}{32\pi^2} \bigg (  \gamma_E - 1 + \log \frac{|m|^2}{4\pi\mu^2}\bigg )+\beta^{\rm b} \,,\qquad\nonumber\\
m_R^{\rm (ren)} &= & -\frac{3 \pi G_N \xi^2 |m|^3}{8\pi^2}\bigg (  \gamma_E + \log \frac{|m|^2}{4\pi\mu^2}\bigg ) +m_R^{\rm b} \,,
\end{eqnarray}
where we denoted with the subscript {\rm b} the unobservable bare constants and included in those the one-loop contribution coming from
renormalization and regularization of the term ${\rm Tr}\log (iS(x;x)[g])$, which leads to Eq.~(\ref{one-loop contribution})~\cite{Davies,Bunch,Christensen}. 
For a more complete discussion on how to obtain the effective action~(\ref{eq2PI.11}), we refer the reader to appendix~\ref{Propagators.Expansion}, 
where we perform perturbative renormalization of the equations of motion and derive and renormalize the 2PI effective action~(\ref{eq2PI.11}). 
Here we limit ourselves to the physical relevance of the result in Eq.~(\ref{eq2PI.11}). To support the physical relevance of our results we show, in the end of appendix~\ref{Propagators.Expansion},
that our perturbative renormalization does not depend on initial conditions for the fermionic field, and we give an example, 
in appendix~\ref{Mode Functions}, that different renormalization schemes leads to the same counter-terms. 
In this section, we have performed renormalization and regularization by constructing an UV expansion of 
the fermionic two-point functions, that is an expansion in inverse powers of $|m|$. The divergent contributions 
to the coincident fermionic two-point function, however, are proportional to the terms $|m| R(x),\, |m|^3,\,\log|m|$,
which appear in this expansion as the only positive powers of the mass. In appendix~\ref{Mode Functions}, 
we perform an opposite expansion, that is $|m|^k$, $k\geq 0$, which is more accurate in the infrared. 
Since quantum divergences are universal, we find in both cases that the same counter-terms renormalize the 2PI effective
action~(\ref{eq2PI.7}).

Since the renormalization scheme used in the appendix~\ref{Propagators.Expansion} is perturbative, the counter terms that we add
in Eq.~(\ref{eq2PI.1}) should be regarded as the first order terms in a series expansion in terms of powers of the coupling constant~$G_N \xi^2$, 
and do not account for higher order contributions. However, we conjecture that higher order corrections do not change the form of~(\ref{eq2PI.11}). 
In other words the renormalized coupling constant in Eq.~(\ref{renormalized coupling constants}) can acquire higher order corrections of~$\mathcal{O} \left(\left ( G_N \xi^2\right )^2\right )$, 
and depend on other geometrical invariants, scalar quantities, as for example $R_{;\mu}R^{;\mu}$,
but our conjecture states we do not have to add counter terms depending on the elements of the Clifford algebra,
as, for example, $R_{;\mu}\bar{\psi}\gamma^{\mu}\psi$.

The effective renormalized action~(\ref{eq2PI.11}) is applicable when the terms in the first line remain perturbative.
A simple order-of-magnitude analysis shows that 
that is the case when 
\begin{equation}
\begin{split}
\label{eq2PI.12} {\rm max}(|m|^2,R)\ll& \frac{M_P^2}{\xi^{2/3}}\,.
\end{split}
\end{equation}
is satisfied which is, for $|\xi|\gg1$, a much higher scale than the na\^ive cut-off scale squared of the theory, 
$M_P^2/\xi^{2}$. The cut-off scale associated with the remaining terms in the effective action~(\ref{eq2PI.11})
is discussed in detail in section~\ref{Validity of the semiclassical treatment}. 
The effective renormalized action~(\ref{eq2PI.11}) is used in the remainder of this paper 
to analyze the evolution of the quantum-corrected fermionic thermal fluid in a contracting universe 
and -- by making use of a self-consistent treatment -- its effect on the evolution of the Universe, also known as the quantum backreaction.


%
%


\section{Analytical solution in the massless regime}
\label{Analytical solution in the massless regime}

Here we begin with our analysis of the theory defined by the renormalized action~(\ref{eq2PI.11}). 
Variation of the action~(\ref{eq2PI.11}), with respect to the metric,
leads to the modified Einstein's equations, which can be written as,
\begin{eqnarray}
\label{renormalized einstein's equations} &&\hskip-0.5cm \left(R_{\mu\nu} \!-\! \frac{1}{2}g_{\mu\nu} R\right )^a =8\pi G_N^{\rm (ren)}\!\Bigg [ -\frac{\Lambda^{\rm (ren)}}{8\pi G_N}g_{\mu\nu} +\alpha^{\rm (ren)} H^{(1)}_{\mu\nu} 
+2 \beta^{\rm (ren)} {\rm Tr}\big(i{\tilde S}^{\rm(reg)}(x;x)\big)  \big ( R_{\mu\nu} \!-\! \frac{1}{2}g_{\mu\nu} R\big) \nonumber  \\
&+&\zeta^{\rm (1), (ren)}H^{(2)}_{\mu\nu}+2 \beta^{\rm (ren)} {\rm Tr}\big (D_\mu D_\nu \big(i{\tilde S}^{\rm(reg)}(x;x)\big) \big )- 2 \beta^{\rm (ren)}g_{\mu\nu} g^{\sigma\rho} {\rm Tr}\big (D_\sigma D_\rho \big(i{\tilde S}^{\rm(reg)}(x;x)\big)\big)
\nonumber\\
&+& {\rm Tr} \bigg( i \gamma_{(\mu} D_{\nu)} (i{\tilde S}^{\rm(reg)}(x;x)) \bigg)- \frac{1}{2}g_{\mu\nu} {\rm Tr} \bigg( \big (i \gamma^{\sigma} D_{\sigma} - m_R - i \gamma^5 m_I\big ) (i{\tilde S}^{\rm(reg)}(x;x))\bigg)\nonumber \\
&+& g_{\mu\nu}\frac{3 \pi G_N \xi^2}{2}\bigg(\!{\rm Tr}\big[i\Pi^{\rm(reg)}(x,x)(i{\tilde S}^{\rm(reg)}(x;x))\big]\bigg)\Bigg ]^{aa}
       \,,
\end{eqnarray}
where~\cite{Davies},
\[\begin{split}
 H^{(1)}_{\mu\nu} =& 2D_\mu D_\nu R - 2 g_{\mu\nu} g^{\sigma\rho} D_\sigma D_\rho R - \frac{1}{2} g_{\mu\nu} R^2 + 2 R R_{\mu\nu}\,, \\
 H^{(2)}_{\mu\nu} =& 2D_\sigma D_{(\nu} R_{\mu)}^\sigma - g^{\sigma\rho} D_\sigma D_\rho R_{\mu\nu} -\frac{1}{2} g_{\mu\nu} g^{\sigma\rho} D_\sigma D_\rho R + 2 R_\mu^\sigma R_{\nu\sigma} - \frac{1}{2}g_{\mu\nu} R_{\sigma\rho}R^{\sigma\rho}
 \,.
 \end{split}\]
In the framework of the Schwinger-Keldysh formalism, the metric field, 
and all the geometrical quantities appearing in the renormalized Einstein's equations~(\ref{renormalized einstein's equations}), 
have indices in polarities space, {\it i.e.} the metric should be ordered along the contour $\mathcal{C}$ as the fermionic propagator is,
and therefore acquires the index $a$. 
This means that the classical {\it off-shell} metric field is split into $g^{\pm}_{\mu\nu}$, which couple respectively to $(i{\tilde S}^{\pm\pm}(x;x))$,
as can be seen from Eq.(\ref{renormalized einstein's equations}): the right hand side has been expressed in matrix notation using
the self-energy matrix defined in~(\ref{self-energy}), and only the diagonal elements, $aa$, are taken and coupled to the metric field $g^{\pm}_{\mu\nu}$. 
The Feynman and Dyson propagators should therefore be inserted in the right hand side of~(\ref{semiclassical Einstein}), 
and the the Einstein's tensor for $g^{\pm}_{\mu\nu}$ is $G^{\pm}_{\mu\nu}=G_{\mu\nu}[g^{\pm}_{\mu\nu}]$.
However, {\it on-shell} the distinction between different polarities of the metric becomes irrelevant, {\it i.e.} $g^{+}_{\mu\nu}=g^{-}_{\mu\nu}$, and therefore 
$G_{\mu\nu}[g^{+}_{\mu\nu}]=G_{\mu\nu}[g^{-}_{\mu\nu}]\equiv G_{\mu\nu}$. 
Therefore the polarities distinction in $T^{\pm}_{\mu\nu}$ really 
does not lead to two independent equations, which can be seen also from the fact that 
the real parts of the Dyson and Feynman propagators, 
in Schwinger-Keldysh formalism, are the same at coincidence.
We therefore choose to take the trace, in polarities space, of Eq.~(\ref{renormalized einstein's equations}) and evaluate the energy momentum tensor,
on the right hand side of Eq.~(\ref{semiclassical Einstein}), using the so-called statistical (Hadamard) 2-points function~\cite{Keldysh,Chou}, 
\begin{eqnarray}
\label{statistical propagator}&&(F(x;x')) = \frac{i}{2}\langle \big[ \bar{\psi}(x), \psi(x')\big] \rangle =\frac{1}{2}\big ( ({\tilde S}^{+-}(x;x')) + ({\tilde S}^{-+}(x;x'))\big )  \\
&& =\frac{1}{2}\big (({\tilde S}^{++}(x;x')) + ({\tilde S}^{--}(x;x'))\big ) \nonumber
=\frac{1}{2} {\rm Tr} \begin{pmatrix} 
(S^{++}(x;x')) & (S^{+-}(x;x')) \\
(S^{-+}(x;x')) & (S^{--}(x;x'))
\end{pmatrix}\,.
\end{eqnarray}
The renormalized equation of motion for the statistical 2-points function is then obtained by summing 
the one-loop corrected renormalized equation for the propagators $iS^{\pm\pm}$~(\ref{2PI EoM: 1 loop:d}),
\begin{eqnarray}
\left (i\cancel{D} \!-\!  m_R^{\rm (ren)} -im_I\gamma^5-\beta^{\rm (ren)} R(x) \right ) (-iF(x;x')) 
   &=&
\label{renormalized EoM statistical propagator}\\
&&\hskip -7.5cm
- 3 \pi G_N \xi^2 \Big\{
   {\rm Tr}\big[(-iF(x;x))\gamma^5\gamma^\sigma\big]
           \gamma^5\gamma_\sigma
   \!-\!\gamma^5\gamma^\sigma(-iF(x;x))\gamma^5\gamma_\sigma\Big\}(-iF(x;x'))
 \nonumber
\,,
\nonumber
\end{eqnarray}
where we have used the fact that the regularized Feyman and Dyson propagators, in the loop contribution to
Eq.~(\ref{renormalized EoM statistical propagator}), are the same since they are evaluated at coincidence. 
Note that such a feature is not general, but follows from the locality of the torsion induced four fermion interaction.

Eqs. (\ref{renormalized einstein's equations}) and (\ref{renormalized EoM statistical propagator}) are the starting point
for analysing a gravitationally induced collapse, and the modification that the torsion interactions induce and constitute the main result of this paper.
We shall assume that all relevant quantities 
appearing in the subsequent equations are renormalized, and hence finite. The details of how this is achieved are 
discussed in section~\ref{2PI effective action renormalization} and appendix~\ref{Propagators.Expansion}, where we use dimensional regularization and renormalization.
We can choose the renormalized parameters, $\Lambda^{\rm (ren)}, \,\alpha^{\rm (ren)},\,\beta^{\rm (ren)}$, appearing in the equation~(\ref{renormalized einstein's equations}) to be small,
such they play no significant role in the evaluation of the fermionic propagator and in the evolution of the universe~(with the exception of $m_R^{\rm (ren)}$). 
We also switch back to $D=4$ in all the integrals and definition, since now leaving those integrals more general serves no purpose. 

Assuming an homogeneous and isotropic universe of the Friedmann-Lema\^ itre-Robertson-Walker (FLRW) type,
for which the metric in conformal coordinates $(x ^0=\eta,x^i)$
reads,
\begin{equation}
  g_{\mu\nu} = a^2(\eta)\eta_{\mu\nu}
\,,\qquad  \eta_{\mu\nu} = {\rm diag}(1,-1,-1,-1)
\,,
\label{metric}
\end{equation}
we can rewrite~(\ref{renormalized EoM statistical propagator}) for the statistical two-point function of the conformally rescaled fields, {\it i.e.} 
\begin{eqnarray*}
\chi &\equiv& a^{3/2} \psi \\
-iF_{\chi}(x;x') &\equiv&\frac{1}{2} \langle\big[ \bar{\chi}(x),\chi(x')\big ]\rangle = -ia(\eta)^3F_{\psi}(x;x')\,,
\end{eqnarray*}
such that the operator $\cancel{D}$ in~(\ref{renormalized EoM statistical propagator}) turns into ordinary derivative~\cite{Prokopec3}.
To avoid a too cumbersome notation, we write $F_{\chi}(x;x')\equiv F(x;x')$, suppressing the 
subscript $\chi$, and $m^{\rm (ren)}_R \equiv m_R$. All the propagators in what follows are assumed to be the ones for the field $\chi$.
We also project the statistical propagator into the helicity basis by means of $P_h$,  $F^{(h)}(x;x') = P_h F(x;x')$, where~\footnote{The 
helicity projector~(\ref{helicity projector}) is correct in $D=4$ only. Its suitable generalization to $D$ dimensions is given in 
{\it e.g.}~\cite{Prokopec3}; since no subsequent results are affected by that generalization, 
for simplicity we quote here the $D=4$ projector.}   
\begin{equation}
P_h \equiv \frac{\mathbb{1} + h  \hat H}{2}
\,,\qquad \hat H\equiv \gamma^0 \hat{k} \cdot \vec{\gamma}\gamma^5
\,,
\label{helicity projector}
\end{equation}
and the $\gamma$ matrices are expressed in Weyl basis. We also assume that, because of spatial homogeneity of the background space-time, 
the fermionic 2 point function assumes the form 
\begin{equation}
\label{Homogeneity ansatz} (-iF(x;x')) = (-iF(\eta,\eta'; \vec{x}-\vec{x}')) \equiv \int\frac{\text{d}\vec{k}}{(2\pi)^3} e^{i \vec{k}\cdot(\vec{x}-\vec{x'})}(-iF(\eta,\eta';\|\vec{k}\|))
\end{equation}

In a FLRW universe the helicity projector commutes with the Dirac operator~\cite{Prokopec3} in (\ref{eq1.10}), 
so that we simply get\footnote{We now drop the notation $m_R^{\rm (ren)}$, and simply write $m_R$, since $m_R^{\rm (ren)}$ is just a parameter determined by experiments.}, 
\begin{eqnarray} 
\label{eq1.11} 
&&\hskip 1cm(i \gamma^0 \partial_\eta -\; hk\: \gamma^0 \gamma^5 -  a (m_R + i \gamma^5  m_I) ) (-iF^{(h)}(\eta,\eta';\vec{k}))  \\
\hskip -1cm&=& -\frac{ 3 \pi G_N \xi^2}{a(\eta)^2}\int\frac{\text{d}\vec{p}}{(2\pi)^3} \Big\{
   {\rm Tr}\big[\sum\limits_{h'}(-iF^{(h')}(\eta,\eta;\vec{p}))\gamma^5\gamma^\sigma\big]
           \gamma^5\gamma_\sigma
   \!-\!\gamma^5\gamma^\sigma(-iF^{(h)}(\eta,\eta;\vec{p}))\gamma^5\gamma_\sigma\Big\}\nonumber\\
  && \times(-iF^{(h)}(\eta,\eta';\vec{k}))
 \nonumber
\,,
\end{eqnarray}
where we have also substituted the operator $\vec{k} \cdot \vec{\gamma} = \|\vec{k}\| \gamma^0 \hat{h} \gamma^5$ and expanded in momentum space. Note that the statement that the helicity operator commutes with the Dirac Hamiltonian is only true in FLRW spacetimes, and it is not the case in less symmetric spaces. It would be hence instructive to study the evolution of fermions in less symmetric collapsing space-times in which helicity is, in general, not conserved. For the scope of the present paper, however, we are going to assume that the symmetry of the initial vacuum state respects that of the background space. In this case  the following block diagonal helicity {\it Ansatz} for the fermionic Hadamard function, in Wigner representation, is exact~\cite{Prokopec2,Prokopec1},
\begin{equation}
\label{eq1.12}
\begin{split}
-iF^{(h)}(k,x)  
\equiv &\frac{1}{2} \int \text{d}^D r e^{i k\cdot r}\sum\limits_{h=\pm}
     {\rm Tr} \left( \hat{\rho}_H \big [\bar{\chi}_{h} (x\!-\!r/2) ,\chi_{h}(x\!+\!r/2) \big ]\right)  \\
=& \sum_{h=\pm} g_{ah}(k,x) \rho^a  \otimes  \frac{1}{4}(\mathbb{1} \!+\! h \hat{k}\cdot \vec{\sigma})
\,,
\end{split}
\end{equation}
where $\sigma^a$ and $\rho^a$ ($a=0,1,2,\dots,D\!-\!1$) are the Pauli matrices ($\rho^0 = \mathbb{1}$) 
and 
\begin{equation}
f_{ah}(\vec k,x)=\int \frac{dk^0}{2\pi}{\rm e}^{ik^0x^0}g_{ah}(k,x)
\,.
\label{fah's}
\end{equation}
Multiplying the Dirac equation (\ref{eq1.11}) and its hermitean conjugate by $\{ \mathbb{1}, \gamma^0, \gamma^5, \gamma^0\gamma^5\}$, respectively, and taking traces over spinorial indices
yields to the semiclassical equations for the currents defined in (\ref{fah's}) 
\begin{subequations}
\begin{eqnarray}
\label{eq1.14.0} \partial_\eta f_{0h}(\vec{k}) &=& 0 , \\
\label{eq1.14.1} \partial_\eta f_{1h}(\vec{k}) +2 h |\vec{k}| f_{2h}(\vec{k}) - 2 a m_I f_{3h}(\vec{k}) &=&\\
=\frac{6 \pi G_N \xi^2}{ a^2} \int \frac{\text{d}\vec{p}}{(2\pi)^3}  \bigg ( \sum_{h_1}f_{3h_1}(\vec{p}) f_{2h}(\vec{k}) &-& \frac{1}{4} \big (f_{3h}(\vec{p}) f_{2h}(\vec{k}) + f_{3h}(\vec{k}) f_{2h}(\vec{p})\big )\bigg ), \nonumber\\
\label{eq1.14.2} \partial_\eta f_{2h}(\vec{k}) - 2 h |\vec{k}| f_{1h}(\vec{k}) + 2 a m_R f_{3h} (\vec{k}) &=&\\
=- \frac{6 \pi G_N \xi^2}{ a^2} \int \frac{\text{d}\vec{p}}{(2\pi)^3} \bigg ( \sum_{h_1}f_{3h_1}(\vec{p}) f_{1h}(\vec{k}) &-& \frac{1}{4} \big (f_{3h}(\vec{p}) f_{1h}(\vec{k}) + f_{3h}(\vec{k}) f_{1h}(\vec{p}) \big )\bigg ), \nonumber \\
\label{eq1.14.3}\partial_\eta f_{3h} -2 a m_R f_{2h} + 2 a m_I f_{1h} &=& 0. 
\end{eqnarray}
\end{subequations}
where we assumed $f_{ah}$ a homogeneous and isotropic state in which case $f_{ah}=f_{ah}(\eta,k)$  $(k=\|\vec k\,\|$)
and we used the Wick theorem to evaluate the interaction terms in the Hartree-Fock approximation~(see Figure~\ref{fig:fig1}), in the same way discussed in appendix~\ref{Propagators.Expansion}.
We have assume that the vacuum contribution has been removed from the currents~(\ref{fah's}) by the 
renormalization and regularization procedure performed in the appendix~\ref{Propagators.Expansion}, such that only 
the regular part of $F$ actually contributes to the equations of motion~(\ref{eq1.14.0}--\ref{eq1.14.3}) and the energy momentum tensor.
In solving the interacting Dirac equation for the currents defined in~(\ref{eq1.14.0}--\ref{eq1.14.3}), 
we will consider the evolution of a fermionic isotropic fluid, 
evolving from an initial thermal state. As we prove in appendix~\ref{Propagators.Expansion} for the two-loop perturbative renormalization of 
the effective action~(\ref{Action and counterterms}), removing the divergent contributions
from an initial Cauchy surface leaves the regular part of the thermal fluid unaffected. 
This is because the regular solution of the equations~(\ref{eq1.14.0}--\ref{eq1.14.0}), from the perspective of the vacuum fluid, 
is just a change of initial conditions, and the renormalization scheme does not depend on initial conditions. 
This statement should hold true at all orders in
the perturbative renormalization of the effective action~(\ref{Action and counterterms}), 
and if we choose to set all the non perturbatively renormalized parameters to be small compared to the Planck mass
Eqs.~(\ref{eq1.14.0}--\ref{eq1.14.0}) are actually the 
correct equations to study particle production in a contracting universe.  

We are now going to solve the Dirac equations~(\ref{eq1.14.0}--\ref{eq1.14.3}) in the light particle limit~{\it i.e.} $|m| \ll k_B T$,
by constructing a perturbation theory in powers of $|m|$ and evaluating the leading order mass correction to Eqs.~(\ref{eq1.14.0}--\ref{eq1.14.3}). 
We choose to study this limit, because it allows us to solve the equations~(\ref{eq1.14.0}--\ref{eq1.14.3}) analytically 
for a general scale factor, as we will see shortly,
and therefore allows to study the backreaction of the fermionic fluid on the geometry numerically. 
Furthermore, the mass of the conformally rescaled fields appears multiplied by the scale factor $a(\eta)$,
in Eqs.~(\ref{eq1.14.0}--\ref{eq1.14.3}), while the temperature scales as $T \propto 1/a(\eta)$,
such that our approximation~($m_{R,I} \ll k_BT$) becomes more accurate during late times of the collapse~($a(\eta)\rightarrow 0$).

Since the torsion induced interaction is local in space, we get momentum integrals in equations (\ref{eq1.14.0} - \ref{eq1.14.3}). To deal with this integro-differential system, we define the momentum integrals of the functions $f_{ah}$. Because of spatial homogeneity, we expect the mode functions to depend only on $k = ||\vec{k}||$. The renormalized functions $f_{ah}(\eta, k)$ for a thermal distribution go to zero faster than any power of $k$, such that the definition
\begin{equation}
\label{eq4.1}
n^{(m)}_{ah}(\eta) = \int\frac{ \text{d}\vec{k}}{(2\pi)^3} k^m f_{ah}(\eta,k),
\end{equation}
is finite for all $m\in\mathbb{N}$. From~(\ref{eq1.14.1}--\ref{eq1.14.3}) we get the following set of equations, 
\begin{subequations}
\begin{eqnarray}
\label{eq4.2.1}\hskip -1cm\partial_\eta n^{(m)}_{h} - 2 h i\; n^{(m+1)}_{h} +  i \frac{ 2\alpha_5}{a^2} \left (  \sum_{h'} n^{(0)}_{3h'} -\frac{n^{(0)}_{3h}}{4} \right )\; n^{(m)}_h - i \left (\frac{\alpha_5 n^{(0)}_{h} }{2a^2} + a m\right ) \! n_{3h}^{(m)} &=& 0,\\
\label{eq4.2.2}\hskip -0.9cm\partial_\eta n^{*(m)}_{h} + 2 h i\; n^{*(m+1)}_{h} -  i \frac{ 2\alpha_5}{a^2} \left (  \sum_{h'} n^{(0)}_{3h'} -\frac{n^{(0)}_{3h}}{4} \right )\; n^{*(m)}_h + i\left (\frac{\alpha_5 n^{(0)*}_{h} }{2a^2}+ a m^*\right ) \!n_{3h}^{(m)} &=& 0,\\
\label{eq4.2.3}\hskip -1cm\partial_\eta n^{(m)}_{3h}  &=& 0.
\end{eqnarray}
\end{subequations}
where we defined $n_h = n_{1h} + i n_{2h}$ and $\alpha_5 =  {3 \pi G_N \xi^2}/{2}$.
From Eq.~(\ref{eq4.2.3})  it follows that $n^{(m)}_{3h}$ is constant, which is the big simplification that 
comes from taking the light particle limit. 
Eqs.~(\ref{eq4.2.1}--\ref{eq4.2.2})  are both complex and moreover they are complex conjugates of each other, such 
that it suffices to solve one of the two.
The moments $n^{(m)}_{ah}$ all couple to each other even in the massless case
because the kinetic term generates moments $\propto n^{(m+1)}$. 
We would like to overcome this problem and diagonalise the system of equations (\ref{eq4.2.1}--\ref{eq4.2.3}). 
This can be achieved by defining a generating function over an auxiliary variable $\rho$ as follows,
\begin{subequations}
\begin{eqnarray}
\label{eq4.3.1} \nu_{h}(\rho) &=& \sum\limits_{m=0}^\infty \frac{n^{(m)}_{h}}{ m!} (\beta\rho)^m, \\
\label{eq4.3.2} \nu_{3h}(\rho) &=& \sum\limits_{m=0}^\infty \frac{n^{(m)}_{3h}}{ m!} (\beta\rho)^m .
\end{eqnarray}
\end{subequations}
where the factor $\beta^m = {1}/{(k_BT_0)^m}$ ($T_0$ is the initial temperature of the fluid)
has been introduced on dimensional grounds. 
With this definition the system of equations (\ref{eq4.2.1}--\ref{eq4.2.3}) simplifies to 
\begin{subequations}
\begin{eqnarray}
\label{eq4.4.2}\hskip 2cm \partial_\eta \nu_{3h}(\rho,\eta) \!=\!0 \!\implies \!\nu_{3h}(\rho,\eta) \!&=&\! \nu_{3h}(\rho), \\
\label{eq4.4.1} \left[\partial_\eta + i\frac{2\alpha_5}{a(\eta)^2}\left(\sum\limits_{h'}\nu_{3h'}(0) -\frac{\nu_{3h'}(0)}{4}\right ) \!-\! \frac{2h i}{\beta}\partial_\rho \right] \nu_h(\rho,\eta) \!\!\!&=&\\
=i \left (\frac{\alpha_5}{2a(\eta)^2} \nu_h(0,\eta)+a(\eta) m \right )\nu_{3h}(\rho)&& \nonumber
\,.\qquad
\end{eqnarray} 
\end{subequations}
Indeed the system of equations (\ref{eq4.4.2} - \ref{eq4.4.1}) is diagonal, but the price we paid for this is that we picked up an extra partial derivative. In other words, our original system of infinitely many equations all coupled to each other is analogous to a system of two first order PDEs. This is not a coincidence, but follows from the structure of (\ref{eq4.2.1}--\ref{eq4.2.3}): written in matrix form, $M \times \vec{n}_h = 0$, one finds that the matrix $M$ is actually a circulant matrix, where each line is obtain from the line above by a constant permutation. Every system satisfying this condition (and some general constraints to assure convergence) can be diagonalised as we do so in this paper. 

The system of equations (\ref{eq4.2.1}--\ref{eq4.2.3}) represents infinitely many differential equations, and therefore it only specifies solutions up to infinitely many integration constants. In our notation, this translates to specify initially the analytical functions 
$\nu_h, \nu_{3h}$, which are given by the initial thermal state. This condition yields
\begin{subequations}
\begin{eqnarray}
\label{eq4.5.1} f_{h} &=& \frac{2m}{\omega_k} \frac{1}{e^{\beta \omega_k } +1}, \\
\label{eq4.5.2} f_{3h} &=& \frac{2hk}{\omega_k} \frac{1}{e^{\beta \omega_k } +1}, \\
\omega_k &=& \sqrt{k^2 + |m|^2} = |k| +\mathcal{O}(m^2), \; m\rightarrow 0\,, \nonumber   
\end{eqnarray}
\end{subequations}
which are obtained from the regular part of the thermal equilibrium propagator~\cite{LeBellac,Prokopec2} and imposing that the initial scale factor, $a(\eta_0)=1$, where $\eta_0$ is some time at which the fields are in thermal equlibrium. 
Note that $f_h$ is actually proportional to $m = m_R + i m_I$. This implies that, when $m_{R,I} \rightarrow 0$, $f_h\rightarrow 0$. This corresponds to the trivial solution, in which the scalar and pseudoscalar currents are zero at all times 
and the pseudo-vector current is initially thermal and, to leading order in $m$, does not depend on the scale factor. 
On the other hand, if one expands all terms in the Dirac equation to $\mathcal{O}(m)$, and neglects $\mathcal{O}(m^2)$, he will get precisely equations (\ref{eq4.4.2}--\ref{eq4.4.1}). This assures that our expansion is 
self-consistent.  

In the light particle limit we can calculate the generating functions, whose leading terms are 
\begin{subequations}
\begin{eqnarray}
\label{eq4.6.1}
\nu_{h}(\rho) &=&\int \frac{\text{d}^3 k}{(2\pi)^3} \frac{2m}{\omega_k} \frac{e^{\beta k\rho}}{e^{\beta \omega_k }\!+\!1}  
= \frac{m (k_B T_0)^2}{2 \pi^2}  \left ( \psi'(1\!-\!\rho) - \frac{1}{2} \psi'\Big(1\!-\!\frac{\rho}{2}\Big) \right )
+\mathcal{O}\big(m^3\big), \\
\label{eq4.6.2}
 \nu_{3h}(\rho) &=&\int \frac{\text{d}^3 k}{(2\pi)^3} \frac{2hk}{a\omega_k} \frac{e^{\beta k\rho}}{e^{\beta \omega_k }\!+\!1}
 = -\frac{h(k_B T_0)^3}{2 \pi^2}  \left ( \psi''(1\!-\! \rho) - \frac{1}{2} \psi''\Big(1\!-\!\frac{\rho}{2}\Big) \right )
 +\mathcal{O}\big(m^3\big)
\,. \quad
\end{eqnarray}
\end{subequations}
The condition $m\ll k_BT_0$ allows us to calculate analytically the initial generating functionals as in (\ref{eq4.6.1}--\ref{eq4.6.2}). Our mass expansion is therefore valid when the mass is small with respect to the initial temperature. Were this not the case for the initial state, the temperature is expected to increase as the universe becomes more closely packed, and the mass terms in the Dirac equation appear multiplied by $a(\eta)$, which implies that their contribution becomes less and less important as $a\rightarrow 0$. The ultra-relativistic limit allows us to calculate the integrals (\ref{eq4.6.1}--\ref{eq4.6.2}) analytically.

To integrate (\ref{eq4.4.2}--\ref{eq4.4.1}) we use the method of characteristic curves~\cite{characteristiccurves}, which in practice means finding a particular solution for some trajectory $(\eta(\lambda), \rho(\lambda))$ and analytically extend the result to all values of the doublet $(\eta, \rho)$. Since $\nu_{3h}$ is constant in time, we only need to solve (\ref{eq4.4.1}). The method of characteristic curves yields to the following solution
\begin{equation}
\begin{split}
\label{eq4.7} 
\nu_h(\rho,\eta) =& \text{exp}\Bigg( \frac{i\alpha_5 \nu_{3h}(0)}{2} \int\limits_{\eta_0}^\eta\text{d}\eta' a(\eta')^{-2}\Bigg)  \\
\times 
\Bigg \{F& \left ( \eta \!+\! \frac{\beta\rho}{2hi}\right )  \!-\! \!
\int\limits_{0}^{\frac{\beta\rho}{2hi}}\! \text{d}s' \bigg[ \frac{i\alpha_5}{2}a\left (\!-s' \!+\! \eta \!+\! \frac{\beta\rho}{2hi}\right )^{-2}  
\!\! F \left(\! -s' \!+\! \eta \!+\! \frac{\beta\rho}{2hi} \right) \nu_{3h}\left( \frac{2hi}{\beta} s'\right) \\
+&a\left (\!-s' \!+\! \eta \!+\! \frac{\beta\rho}{2hi}\right ) im\, \nu_{3h}\left( \frac{2hi}{\beta} s'\right)\bigg ]\!\Bigg \}
\,,
\end{split}
\end{equation}
where $F$ is an analytical function specified by the initial conditions (\ref{eq4.5.1} - \ref{eq4.5.2}). Calculating the energy momentum tensor in terms of the $n^{(m)}$'s one finds that it only depends on $n^{(0)}_h$ and $n^{(0,1)}_{3h}$, which implies that we are interested in the solution $\nu_h(\rho=0,\eta)$.\footnote{Since $n^{(m)}_h=\frac{1}{\beta^m}\frac{\partial^m \nu_h}{\partial \rho^m}\bigg |_{\rho=0}$.} At $\eta = \eta_0$ we want $\nu_h$ to be in the form (\ref{eq4.6.1}), which translates to 
\begin{equation}
\label{eq4.8}
\begin{split}
F( \eta ) \!-\! \int\limits_{\eta_0}^{\eta} \text{d}s \;&   
 \nu_{3h}\left(  \frac{2hi}{\beta} ( \eta\!-\! s ) \right)\left (\frac{i\alpha_5}{2}\frac{F ( s )}{a(s)^2} + a(s)im\right )= \\
& =\frac{m (k_B T_0)^2}{2\pi^2}\left (\psi'\left(1\!-\!\frac{2hi}{\beta}(\eta\!-\!\eta_0)\right) - \frac{1}{2}\psi'\left(1\!-\!\frac{hi}{\beta} (\eta-\eta_0)\right) \right ) 
\,.
\end{split}
\end{equation}
Although this equation is derived using the auxiliary variable $\rho$, it can be rewritten only as a function of conformal time, due to the wave-like nature of (\ref{eq4.4.1}). Eq. (\ref{eq4.8}) is a Volterra integral equation of the second kind~\cite{Yoshida}, which can be solved by iteratively substituting $F$ into the equation~\cite{Hackbusch,Yoshida}. One would then get an infinite series which converges on any finite interval of the real line. Although this approach is viable, this series does not possess good convergence properties. Eventually it converges, but one might have to calculate hundreds of terms before getting close enough to the solution. If one is interested in evolving a fermionic field in a fixed background, however, this approach is good enough, but, since we are interested in the full back reaction, we have to construct some approximation.
\begin{figure}[htbp]
\includegraphics[width=0.8\textwidth]{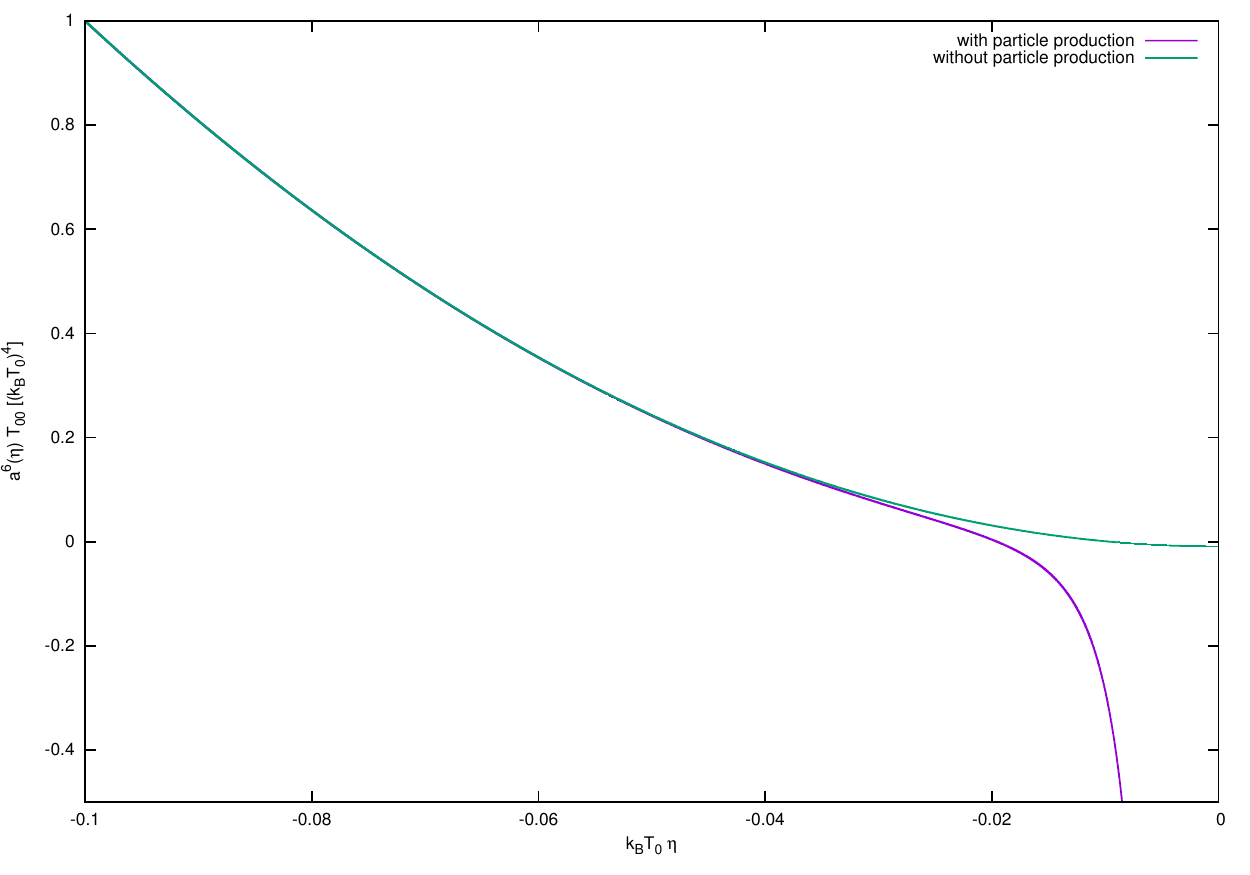}
\caption{The energy density of the conformally rescaled fields. The epsilon parameter, $\epsilon =\frac{1}{a(\eta)}\frac{d}{d\eta}\left ( \frac{1}{H}\right )$, has been set to $2$, its value during radiation domination. The parameters are $\xi \simeq 10^3$, $\frac{|m|}{k_BT_0} \simeq 10^{-2}$ and $k_BT_0 \simeq 0.01M_p$. Note that the tail diverges towards $-\infty$, when $a(\eta) = \left ( \frac{ \eta}{\eta_0}\right )$ becomes small, only if particle production is included.}
\label{fig:fig3}
\end{figure}
In figure \ref{fig:fig3}, we can inspect the evolution of the energy density in a fixed radiation dominated background. We choose to multiply the fluid energy momentum tensor by a factor $\propto a^6(\eta)$, which is the classical scaling of the torsion contributions. Figure \ref{fig:fig3} clearly shows that quantum contribution will induce a faster scaling, meaning that the scaling behaviour will be modified to $a^{6+\gamma_1}(\eta)$, where $\gamma_1 \geq 0$. 
\begin{figure}[htbp]
\centering
\centering
\includegraphics[width=0.7\textwidth]{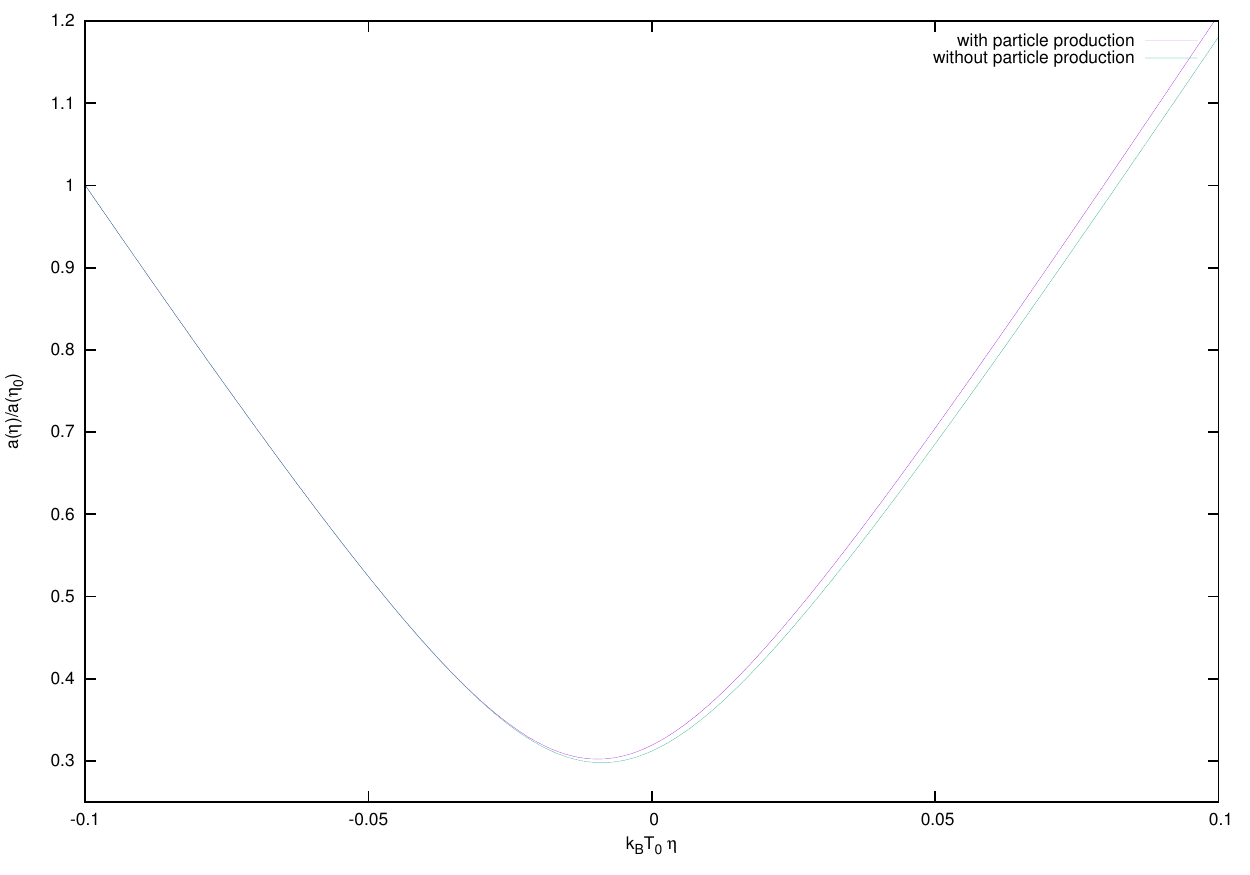}
\centering
\includegraphics[width=0.7\textwidth]{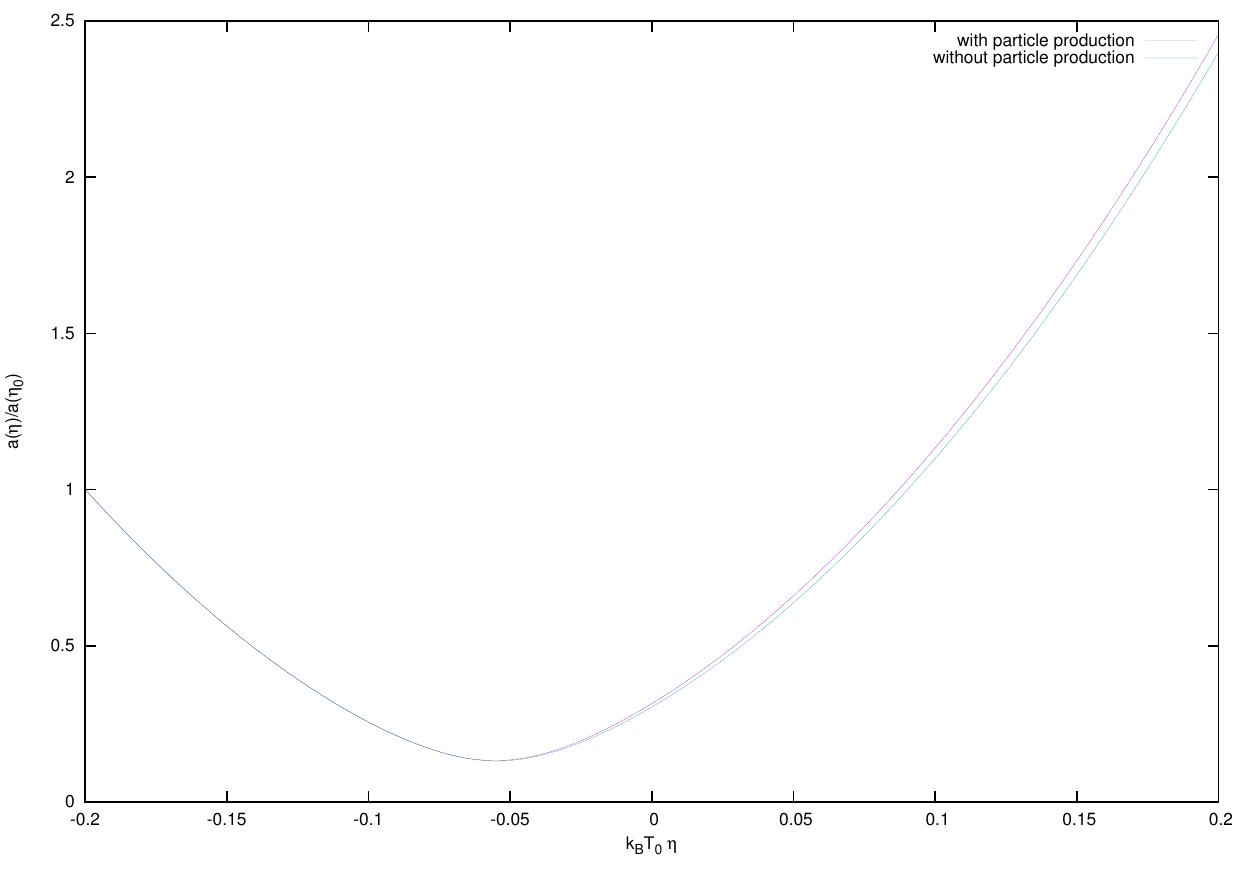}
\caption{Scale factor evolution as a function of conformal time $\eta$, for $\xi \simeq 10^3$, $\frac{|m|}{k_BT_0} \simeq 10^{-2}$ and $k_BT_0 \simeq 0.01M_p$. The evolution starts in a Radiation dominated background, i.e. $a(\eta) =\left (\frac{\eta}{\eta_0}\right)$, in the upper figure, and in matter domination, i.e. $a(\eta) =\frac{1}{2}\left (\frac{\eta}{\eta_0}\right)^2$, in the lower. We notice that the scale factor does not become singular, but it reaches a minimum value $a_\textit{min}$. The effect of particle production is to enhance the bounce, making $R\big |_\textit{bounce}$ slightly smaller. Moreover, introduction of particle production makes the hubble parameter after the bounce bigger: the universe collapses and starts expanding slightly faster than it was collapsing.}
\label{fig:fig4}
\end{figure}


\section{Self-consistent backreaction}
\label{Self-consistent backreaction}

Looking at the structure of (\ref{eq4.8}), and acknowledging that $\alpha_5$ is a rather small parameter, $\mathcal{O}(\xi^210^{-39} GeV^{-2}$), it is clear that torsion contributions are only going to matter at late stages of the collapse. This means that the fermionic field is going to evolve approximately non interacting until the universe gets close to a singularity. We can therefore consider a time $\eta_0$ when the fields are still at thermal equilibrium, and expand Eq.~(\ref{eq4.8}). The asymptotic expansion we are talking about, $\eta \rightarrow 0$, describes the latest phase of the collapse. In this regime (\ref{eq4.8}) can be expanded as
\begin{equation}
\label{eq5.1}
\begin{split}
F( \eta )  - \frac{i\alpha_5}{2} \int\limits_{\eta_0}^{\eta} \text{d}s &\;    \nu_{3h}
\left(\!  - \frac{2hi}{\beta}  s  \right)\frac{F ( s )}{a(s)^2}
\simeq \\
\simeq&\frac{m (k_B T_0)^2}{2\pi^2}\left (\psi'\left(1\!+\!\frac{2hi}{\beta}\eta_0\right) 
         - \frac{1}{2}\psi'\left(1\!+\!\frac{hi}{\beta} \eta_0\right) \right )
\,,
\end{split}
\end{equation}
which can be solved exactly with 
\begin{eqnarray}
\label{eq5.2} 
F( \eta ) &=& F_0 \text{exp}\left (\frac{i\alpha_5}{2} \int\limits_{\eta_0}^{\eta} \frac{\text{d}s}{a(s)^{2}} \nu_{3h}\left(  - \frac{2hi}{\beta}  s  \right)\right ), \\
F_0 &=& \frac{m (k_B T_0)^2}{2\pi^2}\left (\psi'\left(1+\frac{2hi}{\beta}\eta_0\right) - \frac{1}{2}\psi'\left(1+\frac{hi}{\beta} \eta_0\right) \right ) 
\,,
 \nonumber
\end{eqnarray}
as it can be easily confirmed by differentiating~(\ref{eq5.1}). Note that we have also dropped the mass term from equation~(\ref{eq4.7}) in the asymptotic expansion~(\ref{eq5.1}). The reason for this is that the mass of the conformally rescaled fields is $\propto a(\eta)$, while the torsion contributions scale $\propto a(\eta)^{-2}$. Therefore, the mass term does not lead to any substantial particle production in the late phase of the collapse, while the torsion contribution does.

\begin{figure}[htbp]
\centering
\includegraphics[width=0.8\textwidth]{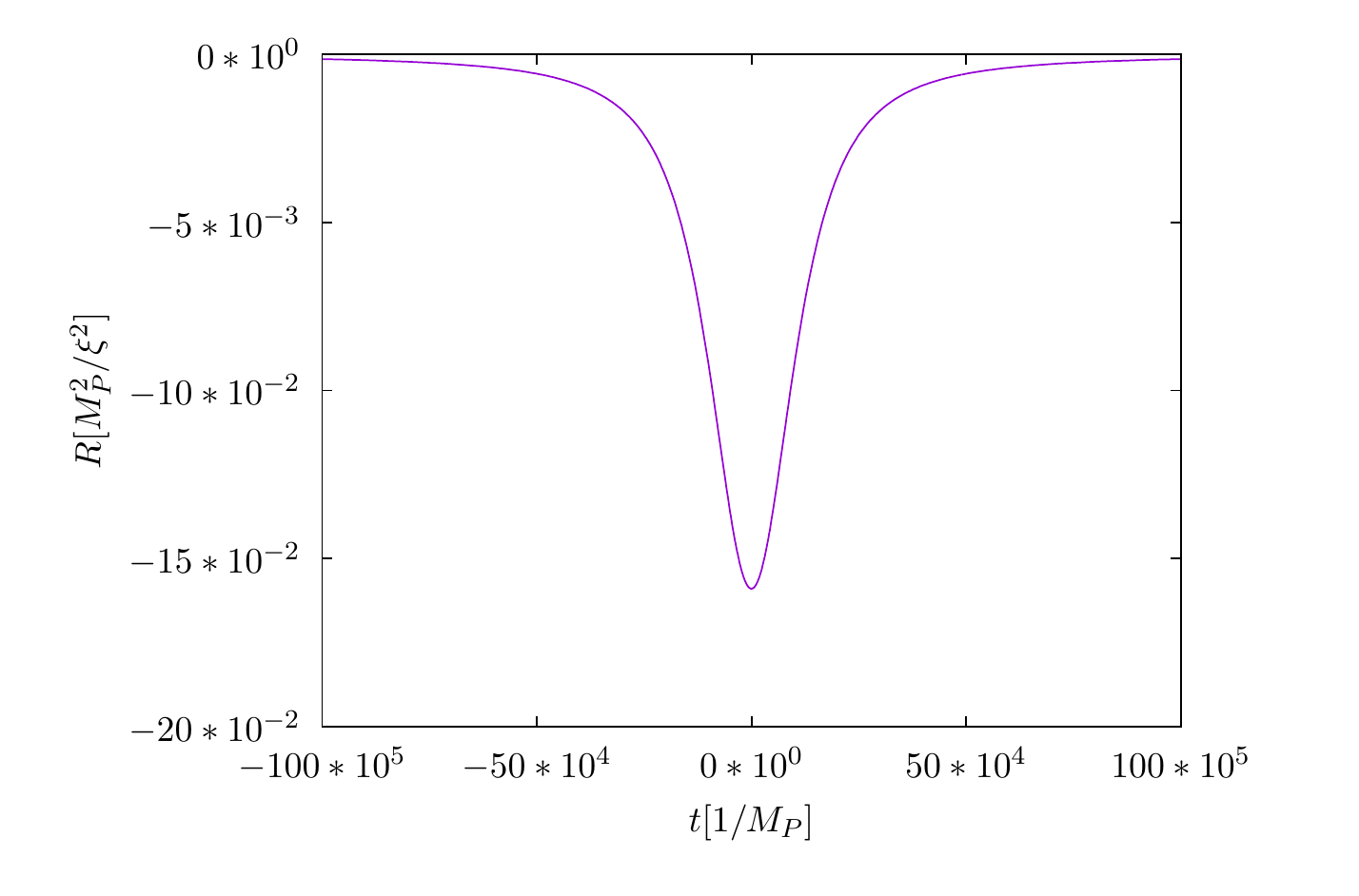}
\caption{Ricci scalar for the matter dominated collapse in units of the unitarity breaking bound, $M_P^2/\xi^2$ as a function of time in Planck units, with $\xi \simeq 10^3$, $\frac{|m|}{k_BT_0} \simeq 10^{2}$ and $k_BT_0 \simeq 0.001M_p$. We infer that the Ricci scale reached at the bounce is substantially lower than the unitarity breaking bound, about ten times smaller. This shows that we can trust the result of our semi-classical approximation, for the matter dominated bounce, since the curvature scale~(as well as the temperature scale) will remain lower than the unitarity breaking bound. }
\label{fig:fig5}
\end{figure}

We can express the energy-momentum tensor in terms of $F$ starting from the energy-momentum tensor (\ref{renormalized einstein's equations}), using the Dirac equation and our Ansatz for the Hadamard function~(\ref{eq1.12}):
\begin{eqnarray}
\label{eq5.3.1} 
\langle T_{00} \rangle&=& \sum_h \int\frac{\text{d}\vec{p}}{(2\pi)^3} \left ( \frac{ 1}{a^4} h |\vec{p}| f_{3h} + \frac{1}{a^3} (m_R f_{1h} + m_I f_{2h} )\right )- \\
&&-\frac{\alpha_5}{a^6}  \int \frac{\text{d}\vec{p}}{(2\pi)^3} \frac{\text{d}\vec{p'}}{(2\pi)^3} \left (\sum_{hh'} f_{3h} f_{3h'} +\sum_h  \frac{1}{2}(f^2_{3h} + f^2_{0h}) + (f_{1h}^2 + f_{2h}^2)  \right ), \nonumber \\
\label{eq5.3.2} 
\langle T_{ij}\rangle &=& \delta_{ij} \bigg \{  \sum_h \int\frac{\text{d}\vec{p}}{(2\pi)^3}  \frac{1}{3}\bigg ( \frac{1}{a^4} h |\vec{p}| f_{3h} \bigg ) - \\
&&-\frac{\alpha_5}{a^6} \sum_{hh'} \int \frac{\text{d}\vec{p}}{(2\pi)^3} \frac{\text{d}\vec{p'}}{(2\pi)^3} \left (\sum_{hh'} f_{3h} f_{3h'} +\sum_h \frac{1}{2}(f^2_{3h} + f^2_{0h}) +  (f_{1h}^2 + f_{2h}^2)  \right )
 \bigg \}
\,,
\nonumber 
\end{eqnarray}
which in the light particle approximation and using the definitions can be simplified to,
\begin{eqnarray}
\label{eq5.4.1} 
\langle T_{00}\rangle &=& \sum_h  \left ( \frac{ 1}{a^4} h \frac{1}{\beta} \frac{\partial \nu_{3h}}{\partial \rho}\right )\bigg |_{\rho=0}-\frac{\alpha_5}{a^6}   \left (\sum_h  \frac{1}{2}(\nu^2_{3h} +2  \nu_h^* \nu_h )  \right )\bigg |_{\rho=0},  \\
\label{eq5.4.2} 
\langle T_{ij}\rangle &=& \delta_{ij} \bigg \{  \sum_h   \frac{1}{3}\bigg ( \frac{h}{a^4} \frac{1}{\beta} \frac{\partial \nu_{3h}}{\partial \rho} \bigg )\bigg |_{\rho=0}-\frac{\alpha_5}{a^6}  \left (\sum_h  \frac{1}{2}(\nu^2_{3h} +2  \nu_h^* \nu_h )  \right )\bigg |_{\rho=0}\bigg \}\,,
\end{eqnarray}
where we have also removed the divergent part of the energy momentum tensor, as described in appendix~\ref{Propagators.Expansion}. 
If one assumes the renormalization parameters rescaled with the appropriate power of the Planck mass to be of $\mathcal{O}(1)$,
corrections to Eqs.~(\ref{eq5.4.1}--\ref{eq5.4.2}), which we reported in Eq.~(\ref{renormalized einstein's equations}), 
are only going to be important at the energy scale of $\mathcal{O}\left (M_P^2/\xi^{2}\right)$,
which is much bigger than the energy scale reached at the moment of the bounce~(see section~\ref{Validity of the semiclassical treatment} and figure~\ref{fig:fig5} for more details), 
and have therefore been neglected in writing Eqs.~(\ref{eq5.4.1}--\ref{eq5.4.2}).
In Figure \ref{fig:fig3} we report  the evolution, in a fixed background, of the rescaled energy 
density~\footnote{We choose to rescale the zeroth component of the energy-momentum tensor 
with the highest power of the scale factor found in it, 
the torsion contributions scaling $\propto a^{-6}(\eta)$.},
\[a^6(\eta) T^\psi_{00}.\] 
Note that the contribution of torsion makes $T^\psi_{00}$ negative, and the particle production acts to enhance this effect at late time. Since this evolution is obtained by keeping $\epsilon\equiv-\dot{H}/H^2$ constant, it does not reproduce a physical behaviour: in reality, $T^\psi_{00}$ appears on the right hand side of the Friedmann equations, in such a way that the sum of the fermionic contribution and the other fluids composing the universe remains positive. In the realistic situation, the total energy density will never become negative, but $\epsilon$ will adjust itself accordingly.

Equations (\ref{eq5.4.1}-\ref{eq5.4.2}) can be substituted in the second Friedmann equation, which in conformal time reads
\[\frac{a''}{a^3} = \frac{4 \pi G_N}{3} (\rho - 3 p). \]
Since we are considering a collapsing FLRW universe, we need to specify which fluid drives the collapse, 
{\it i.e.} we need to specify the background energy density and pressure, which we denotes as $\rho_b$, $p_b$. 
Splitting the background contribution~(which contain also the matter and radiation components of the fermionic fluid) 
from the contribution induced purely by torsion, yields to,
\begin{equation}
\label{eq5.5} \frac{a''}{a^3} = \frac{4 \pi G_N}{3}(\rho_b - 3 p_b) + \frac{4 \pi G_N}{3} \frac{2\alpha_5}{a^6} \sum_h \left ( \frac{1}{2}\nu^2_{3h}|_{\rho=0} + \big | F_h(\eta)\big |^2\right )
\,,
\end{equation}
where $F_h(\eta)$ is given by (\ref{eq5.2}). Once the background evolution is given ({\it i.e.} $w_b(\eta)=p_b/\rho_b$ is known), Eq.~(\ref{eq5.5}) becomes the equation for the scale factor $a(\eta)$ and can be solved self-consistently. Given initial conditions $a_0$, $H_0$ and $k_BT_0$ we can evolve in the background until the second term in (\ref{eq5.5}) becomes significant and then write a numerical code that construct the remainder of the solution. Again, since the torsion contributions are going to be important at late times, we can use the expansion (\ref{eq5.2}) around $k_BT_0 \eta \simeq 0$. 

The results of this numerical analysis are shown in Figure \ref{fig:fig4}. The effect of including particle production, is that the bounce gets enhanced and happens sooner. The magnitude of this effect depends on the initial conditions, but the rest of our conclusions are general. These examples illustrate the importance of taking the backreaction self-consistenly into account for the comparison of a classical evolution with a quantum evolution where perturbative loop effects are self-consistently accounted for.


\section{Validity of the semiclassical treatment}
\label{Validity of the semiclassical treatment}

In this section we address the question of validity of the semiclassical treatment. 
A particular emphasis is on the higher order loop corrections close to 
the bounce, where the curvature scalars reach their maximum. Before we begin our discussion, we observe that 
the coupling $\xi$ (and $\xi'$) can change the perturbatiove cutoff scale, which is in general relativity given by the Planck scale,
to a scale $E_{\rm cutoff}$ which, in the case when $\xi^2\gg 1$, is below the Planck scale.  In order to estimate the new 
cutoff scale of our effective 
theory, consider 2-to-2 fermionic scatterings induced by our four-fermion interaction~(\ref{eq1.8}). The corresponding transition amplitude
is of the order, ${\cal A}_{2\rightarrow 2}\sim k^2G_N\xi^2$. Unitarity is violated when the tree scattering amplitude becomes 
of the order of unity, ${\cal A}_{2\rightarrow 2}\sim 1$, and that happens for incoming particle momenta and energies of the order,
$E,\|\vec k\|\sim M_{\rm P}/\xi\equiv E_{\rm cutoff}$. Since in a relativistic plasma, $\langle E\rangle\sim k_BT$, 
the unitarity bound for the background fluid at the bounce $(b)$ is given by, 
$\rho_b\sim g_t (k_BT_b)^4\ll g_t E_{\rm cutoff}^4\sim M_{\rm P}^4/\xi^4$, where $g_t $ denotes 
the total number of relativistic degrees of freedom. When fermions are light, $|m|\ll k_BT_b$, then 
$E_{\rm cutoff}\sim k_{\rm cutoff}$ and when they are 
heavy, $|m|\gg k_BT_b$, then  $E_{\rm cutoff}\simeq |m|$. 

From Eqs.~(\ref{eq5.3.1}--\ref{eq5.3.2}) we see that the modified semiclassical Friedmann equation 
can be schematically written as, 
\begin{equation}
H^2 \sim \frac{1}{M_{\rm P}^2} \left ( \rho - \frac{\xi^2}{M_{\rm P}^2} n^2 \right )\overset{bounce}{\sim}
\;\;0
\,, 
\label{validity:1}
\end{equation}
where $\rho$ denotes the background energy density (which, close to the bounce, can be safely assumed to exhibit radiation scaling,
$\rho \propto g_t T ^4 \propto 1/a(t)^4$) and $n$ stands for either scalar, pseudoscalar or axial-vector density. 
In a relativistic fluid, $n\sim g_f (k_BT)^3$ ($n\sim g_f|m|(k_BT)^2$) for the axial-vector ((pseudo-)scalar) density.
From~(\ref{validity:1}) it follows that maximum density reached at the bounce is, 
\begin{equation}
 n_b\sim  \frac{M_{\rm P}}{|\xi|} \rho_b^{1/2} 
\,.
\label{validity:2}
\end{equation}
 In relativistic regime  the axial-vector current scales as $n_b\sim g_f (k_BT_b)^3\sim (g_f /g_t^{3/4} )\rho_t^{3/4} $, while 
the (pseudo-)scalar density scales as $n_b\sim g_f |m|(k_BT_b)^2\sim  |m| (g_f /g_t^{1/2} )\rho_t^{1/2} $, 
and Eq.~(\ref{validity:2}) yields
\begin{equation}
  \rho_b^{1/4}\sim  \frac{M_{\rm P}}{|\xi|}\frac{g_t^{3/4}}{g_f}  
\,,\qquad
 |m| \sim  \frac{M_{\rm P}}{|\xi|} \frac{g_t^{1/2}}{g_f} 
\,.
\label{validity:2b}
\end{equation}
The former relation represents a more stringent relation in the relativistic case, in which the UV unitarity bound is 
$E_{\rm cutoff}\sim M_{\rm P}/|\xi|$, while the latter relation represents a more stringent bound in the non-relativistic case,
in which the UV unitarity bound is at $E_{\rm cutoff}\simeq |m|\sim M_{\rm P}/|\xi|$. This means that in the relativistic cases close to equilibrium, Eq.~(\ref{validity:2b}) implies that the bounce occurs at the energy/momentum 
scale above which unitarity is broken, such that in that case the semiclassical approximation used here is not reliable. In the non relativistic case, however, it is still possible to avoid reaching the unitarity bound, if the fermions masses satisfy $m_f \ll \frac{M_P}{|\xi|}$.

 This does not mean that one cannot use semiclassical approximation to study bounce reliably. In what follows we argue that 
semiclassical approximation yields quantitatively reliable results when heavy fermions that are not 
in thermal equilibrium dominate the bounce. Such a situation can occur, for example, when a large star is collapsing.
The simplest possible deviation from equilibrium can be modeled by a chemical potential $\mu$, such that particles are modeled by 
a positive chemical potential and antiparticles by a chemical potential that is equal in magnitude but negative in sign.
Furthermore, for most of non-relativisistic situations, it is a good approximation to assume that $\mu\simeq|m|$.
In that case the pseudo-vector ((pseudo-)scalar) particle density can be approximated by $n\sim |m|^2 k_BT$
($n\sim (|m|k_BT)^{3/2}$). Relative to these, the corresponding antiparticle densities are suppressed 
by a factor, ${\rm e}^{-2|m|/(k_BT)}\ll 1$, and can thus be neglected.
Inserting these into~(\ref{validity:2}) and taking account of the condition $k_BT_b\ll |m|\ll m_{\rm P}/\xi$, we get,
\begin{eqnarray} 
(k_BT_b)^3&\ll&\mu^3
   \simeq |m|^3\sim \frac{m_{\rm P}^2}{\xi^2}\frac{g_t}{g_f^2}(k_BT_b)\ll  \frac{m_{\rm P}^3}{\xi^3}\frac{g_t}{g_f^2}
\,,\nonumber\\
(k_BT_b)^2&\ll&\mu^2
 \simeq |m|^2\sim \frac{m_{\rm P}}{\xi}\frac{g_t^{1/2}}{g_f}(k_BT_b)\ll  \frac{m_{\rm P}^2}{\xi^2}\frac{g_t^{1/2}}{g_f}
\,
\label{valid!}
\end{eqnarray}
for the axial-vector and (pseudo-)scalar densities, respectively. We see that in all cases the temperature at the bounce is much below 
the unitarity scale, $E_{\rm cutoff}\sim m_{\rm P}/\xi$, 
rendering the semiclassical treatment realiable.
As a last comment, it was already noticed~\cite{Hehl1} that in general the bounce occurs at densities, $n_b\sim 1/\ell_b^3$, 
that generate a new length scale $\ell_b$, where  
\begin{equation}
      \frac{\ell_b}{\ell_{\rm P}}\sim \frac{\xi^{1/3}}{(\ell_{\rm P}^4\rho_b)^{1/6}}
\simeq \frac{\xi^{1/3}}{(\ell_{\rm P}^2R_b)^{1/6}}
\,,
\label{new scale}
\end{equation} 
where $R_b$ denotes the Ricci scalar at the bounce.
The length scale $\ell_b$ corresponds to a typical wavelength of sound waves created at the bounce and,
 as the above analysis shows, $\ell_b$
can be significantly larger than the Planck length. This implies that torsion can screen singularities on length scales that 
are significantly larger than the Planck scale, at which the effects of quantum gravity become non-perturbatively large.
Consequently, by making use of perturbative methods one can reliably study the quantum effects 
of matter and gravitational fields close to and at the bounce in Cartan-Einstein theory.

 In conclusion, a detailed analysis of the semiclassical approximation utilised in this work shows that it gives reliable results
for non-relativistic out-of equilbrium fermionic fluids, while in the case when the fermions are relativistic 
the bounce occurs at the scale at which the theory violates the tree-level unitarity, implying that in that case the semiclassical treatment 
in general does not yield reliable results. This, of course, does not mean that there is no bounce in the latter case; it just means
that bounce cannot be reliably established by semiclassical methods.


\section{Conclusion and discussion}
\label{Conclusion and discussion}

We renormalize the 2PI effective action for Einstein-Cartan-Sciama-Kibble theory, at two-loop expansion, 
in {\it in-in} formalism, which leads to the one-loop renormalized equations of motion~(\ref{renormalized EoM statistical propagator})
and the two-loop renormalized energy-momentum tensor~(\ref{renormalized einstein's equations}). 
Our procedure is novel when compared to standard references~\cite{Davies,Bunch,Christensen} on the subject, 
where this procedure is done in {\it in-out} formalism. However, we find the same expression for the 
one-loop effective action, {\it i.e.} ${\rm Tr}\log (iS_0(x;x)[g])$, as seen in the 2PI effective action Eq.~(\ref{2PI effective action, onshell expression}),
which shows that, at least for the free theory, the two formalisms agree. We then perturbatively renormalize the one-loop Dirac equation~(\ref{2PI EoM: 1 loop}), in a general curved 
background, and find that, in order to remove divergencies, we need to introduce mass renormalization, 
a cosmological constant~(with negative sign), $\sqrt{-g}R^2$, Newton constant renormalization and 
a novel term, $\sqrt{-g} \bar{\psi}\psi R$, which acts as a space-time dependent mass for fermions. 

In section~\ref{Analytical solution in the massless regime}, we analyze the effect of torsion of the Einstein-Cartan theory on the evolution of the Universe.
In particular, we study the torsion contribution on a matter and radiation dominated collapsing universe
and find that -- instead of ending in a big crunch singularity -- the universe undergoes a bounce. We have evidence
that this behavior is generic, and is not affected by the nature of the collapse.
In contrast to older works~\cite{Poplawski2,Poplawski1,Trautman},
we do not assume a classical form of the spin fluid sourcing torsion, but instead we derive our description from a {\it full
microscopic treatment of fermions}. We do both: (a) a classical treatment (in which the fermionic fluid is described by
an initial thermal state and particle production due to Universe's  contraction is switched off) and (b) perturbative quantum
treatment (in which particle production is accounted for at the one-loop level in the fermionic dynamical equations
and at the two-loop level as quantum backreaction in the Friedmann equation). Our analysis shows that, for typical initial conditions,
the classical bounce is not significantly affected by
the quantum particle production. We find that, when the fermion production is taken account of, then
the bounce occurs somewhat earlier, indicating that fermion production induces a negative backreaction on the Universe's
evolution, as can be clearly seen from figure~\ref{fig:fig3}. 
This result is actually rather interesting from the perspective of the bouncing cosmologies models. In fact, one of the 
problems of this models is the scaling of cosmological perturbation: during matter dominated collapse, the growing mode of cosmological perturbations 
scale as~$\propto 1/a(\eta)^{11/2}$~\cite{Mukhanov}, which is slower than the torsion contributions. Therefore, 
perturbations get more and more diluted as the universe collapses, yielding to the isotropic universe we observe today. 
During a radiation dominated collapse, however, perturbations scale as~$\propto 1/a(\eta)^{6}$~\cite{Mukhanov}, 
so they would lead to an inhomogeneous universe after the collapse. However, by including particle production, 
we can infer that the torsion contributions actually scale faster than perturbations during radiation domination, 
which makes the model still viable.

The reason why we chose a matter dominated collapse is that the resulting bounce might present a viable alternative to inflation.
Namely, it is well known that the (Bunch-Davies) vacuum state in matter era yields a
flat spectrum of perturbations.~\footnote{This is so because the equation of motion for a conformally rescaled massless scalar,
$\phi_c = a\phi$) in momentum space and in conformal time, $(\partial_\eta^2 + k^2-a^{\prime\prime}/a)(a\phi(\eta,k))=0$,
is identical to that in de Sitter space. This follows immediately from the form of the scale factors,
which is in de Sitter space, $a\propto-1/\eta$, and in matter era, $a \propto \eta^2$, such that in both cases
$a^{\prime\prime}/a=2/\eta^2$. Since the spectrum of a massless scalar in
a Bunch-Davies vacuum in de Sitter space is scale invariant, so must be
the corresponding spectrum of a massless scalar in matter era.}
Therefore,  it would be of a particular interest to derive the power spectrum of cosmological perturbations and
investigate whethere it can be used to seed the large scale structure and fluctuations in cosmic microwave background
that match the data.

A second situation in our study can be of use is that of a collapsing star turning into a black hole.
In this case the interior of the star can be modelled by a FLRW metric~\cite{Mersini},
as long as the collapse respects spherical symmetry.
Therefore, at least in the bulk of the star, the analysis of this paper applies,
and can be used to infer black hole formation. As it happens with the singularity at the beginning of our Universe,
it is probable that also formation of black hole singularities is prevented by torsion. However, that has not yet been 
demonstrated rigorously. We plan to address this question in forthcoming works, using the framework developed in this paper.


\appendix
\section{When torsion couples to the vector current}
\label{appendix A}

  Here we study the case when torsion couples to the  fermionic vector current, {\it i.e.} when 
$\xi=0$ and $\xi'\neq 0$. In particular we shall derive the equations of motion for the fermionic currents
and the corresponding energy momentum tensor and compare with the case when 
torsion couples to the axial vector current discussed in the main text of the paper.
 In the case when $\xi=0$ and $\xi'\neq 0$ the Dirac equation reads,
\begin{equation}
\label{eqA.1} 
\left (i \gamma^\mu D_\mu   - m_R - i m_I \gamma^5 \right)\psi = -(3 \pi G  \xi^2) (\bar{\psi}  \gamma^\sigma \psi) \gamma_\sigma \psi 
\,,
\end{equation}
which in semi-classical approximation, for the currents $f_{ah}$ becomes
\begin{subequations}
\begin{eqnarray}
\label{eqA.2.0} \partial_\eta f_{0h}(\vec{k}) &=& 0 , \\
\label{eqA.2.1} \partial_\eta f_{1h}(\vec{k}) +2 h |\vec{k}| f_{2h}(\vec{k}) - 2 a m_I f_{3h}(\vec{k}) &&\\
=-\frac{18 \pi G_N {\xi'}^2}{ a^2} \int \frac{\text{d}\vec{p}}{(2\pi)^3}  \bigg ( \big (f_{3h}(\vec{p}) f_{2h}(\vec{k}) + f_{3h}(\vec{k}) f_{2h}(\vec{p})\big )\bigg ), \nonumber\\
\label{eqA.2.2} \partial_\eta f_{2h}(\vec{k}) - 2 h |\vec{k}| f_{1h}(\vec{k}) + 2 a m_R f_{3h} (\vec{k}) &&\\
= \frac{18 \pi G_N {\xi'}^2}{ a^2} \int \frac{\text{d}\vec{p}}{(2\pi)^3} \bigg ( \big (f_{3h}(\vec{p}) f_{1h}(\vec{k}) + f_{3h}(\vec{k}) f_{1h}(\vec{p}) \big )\bigg ), \nonumber \\
\label{eqA.2.3}\partial_\eta f_{3h} -2 a m_R f_{2h} + 2 a m_I f_{1h} &=& 0. 
\end{eqnarray}
\end{subequations}
Note the absence of terms containing $\sum\limits_h$, which in the first part of this paper was mainly due to the presence of the $\gamma^5$ matrix in the interaction term, and the consequent violation of parity symmetry. Aside from this and a factor of 3 difference in the torsional coupling constant, the structure of these equations is precisely the same of Eq. (\ref{eq1.14.0}--\ref{eq1.14.3}). In both cases the torsion interactions induce a shift in the mass and in the momenta of the fermionic fields. This effect in case $\xi=0$ is due to the Fock contraction, while in case $\xi'=0$ to both Hartree and Fock. This suggests that both interaction terms can be treated in the same way. Some interest could come from the mixed situation, in which $\xi\, ,\xi' \neq 0$, in which case we would find Eq. (\ref{eq1.14.0}--\ref{eq1.14.3}) again, 
but now the Hartree and Fock terms will have a different coupling strength (respectively, $\xi^2$ and $\xi^2 + 3 {\xi'}^2$). 

The energy momentum tensor reads,
\begin{eqnarray}
\label{eqA.3.1} 
\langle T_{00}\rangle &=& \sum_h \int\frac{\text{d}\vec{p}}{(2\pi)^3} \left ( \frac{ 1}{a^4} h |\vec{p}| f_{3h} + \frac{1}{a^3} (m_R f_{1h} + m_I f_{2h} )\right )- \\
&&-\frac{\alpha_v}{a^6}  \int \frac{\text{d}\vec{p}}{(2\pi)^3} \frac{\text{d}\vec{p'}}{(2\pi)^3} \left (\sum_{hh'} f_{0h} f_{0h'} +\sum_h(f_{1h}^2 + f_{2h}^2) -\frac{1}{2}(f^2_{3h} + f^2_{0h})   \right ), \nonumber \\
\label{eqA.3.2} 
\langle T_{ij}\rangle &=& \delta_{ij} \bigg \{  \sum_h \int\frac{\text{d}\vec{p}}{(2\pi)^3}  \frac{1}{3}\bigg ( \frac{1}{a^4} h |\vec{p}| f_{3h} \bigg ) - \\
&&-\frac{\alpha_v}{a^6} \sum_{hh'} \int \frac{\text{d}\vec{p}}{(2\pi)^3} \frac{\text{d}\vec{p'}}{(2\pi)^3} \left (\sum_{hh'} f_{0h} f_{0h'} +\sum_h (f_{1h}^2 + f_{2h}^2) - \frac{1}{2}(f^2_{3h} + f^2_{0h})   \right )
 \bigg \}
\,,
\nonumber 
\end{eqnarray}
where now $\alpha_v = 3\pi G_N {\xi'}^2/2$. Again there is a structural difference between the vector interaction and the pseudo-vector interaction. Most notably, in the vector case, the term $\propto \sum\limits_h f_{3h}^2$ comes with the opposite sign, when compared to the pseudo-vector case. This will have the effect of delaying the bounce. In this case the classical solution will only lead to a bounce if $|m| > k_B T$, however particle production might prevent the singularity formation in this case too, since $f_{1h}$ and $f_{2h}$ scale faster than $f_{3h}$, which remains constant in the massless regime.  To make a more quantitative statement, one could have to solve 
self-consistently the corresponding semiclassical Einstein equations together with the equations of 
motion~(\ref{eqA.2.0}--\ref{eqA.2.3}), which is beyond the scope of this paper.


\section{Perturbative renormalization} 
\label{Propagators.Expansion}

In this appendix we show how to derive and regularize the equations of motion and the effective 2PI action mentioned in 
section~\ref{2PI effective action renormalization}, from which one can obtain the renormalized stress-energy tensor
that is used in the semiclassical Friedmann equations. 

We start this procedure from the 2PI effective action at two-loop~(\ref{eq2PI.7}--\ref{2PI action: 2 loop}), 
to which we add the local counter terms~(\ref{eq2PI.1}) needed to renormalize the equations of motion, 
\begin{eqnarray}
&&\Gamma[S, g_{\mu\nu}] +S^{(\text{ct})}[ S, g_{\mu\nu}] = \nonumber \\
&=& \int\!\text{d}^D x \sqrt{-g}\sigma_z \left [
   \frac{\Delta\Lambda}{16\pi G_N}\! -\! \frac{\Delta G_N^{-1}}{16\pi } R\! +\! \Delta\alpha R^2 \!+\! \Delta\zeta R_{\mu\nu}R^{\mu\nu}
       \!-\! \Big[ \Delta m_R\!+\!\Delta\beta R\Big] \big(iS(x;x)\big) 
     \right ]
 \nonumber\\
      &&+\int\!\text{d}^D x \sqrt{-g}\!\Big[\sigma_z\text{Tr}[ \left (i\cancel{D}\! -\! m_R\! -\! im_I\gamma^5 \right ) (iS(x;x))]\!-\! i \text{Tr}\log  (iS(x;x))\Big] \nonumber\\
      &&+ \frac{3 \pi G_N \xi^2}{2}\sum_{a,b=\pm}\!\int\!\text{d}^D x \sqrt{-g}
   \Big[\sigma_z^{ab}{\rm Tr}\big[(iS^{bb}(x;x))\gamma^5\gamma^\sigma\big]
       {\rm Tr}\big[(iS^{bb}(x;x))\gamma^5\gamma_\sigma\big]
\nonumber\\
       &&\hskip 3.99cm
   -\sigma_z^{ab} {\rm Tr}\big[(iS^{bb}(x;x))\gamma^5\gamma^\sigma(iS^{bb}(x;x))\gamma^5\gamma_\sigma\big]\Big]
\,.
\qquad\label{Action and counterterms}
\end{eqnarray}
It will become clear in the following how the counter terms  
in~(\ref{Action and counterterms}) contribute to the equations of motion, and how they cancel the divergent parts. 
Also note the extra, purely gravitational counter-terms:
those do not contribute to the equations of motion for the fermionic two-point function, 
but they are required to renormalize the effective action and $\langle T_{\mu\nu}\rangle$.

\begin{figure}
\centering
    \includegraphics[width=5in]{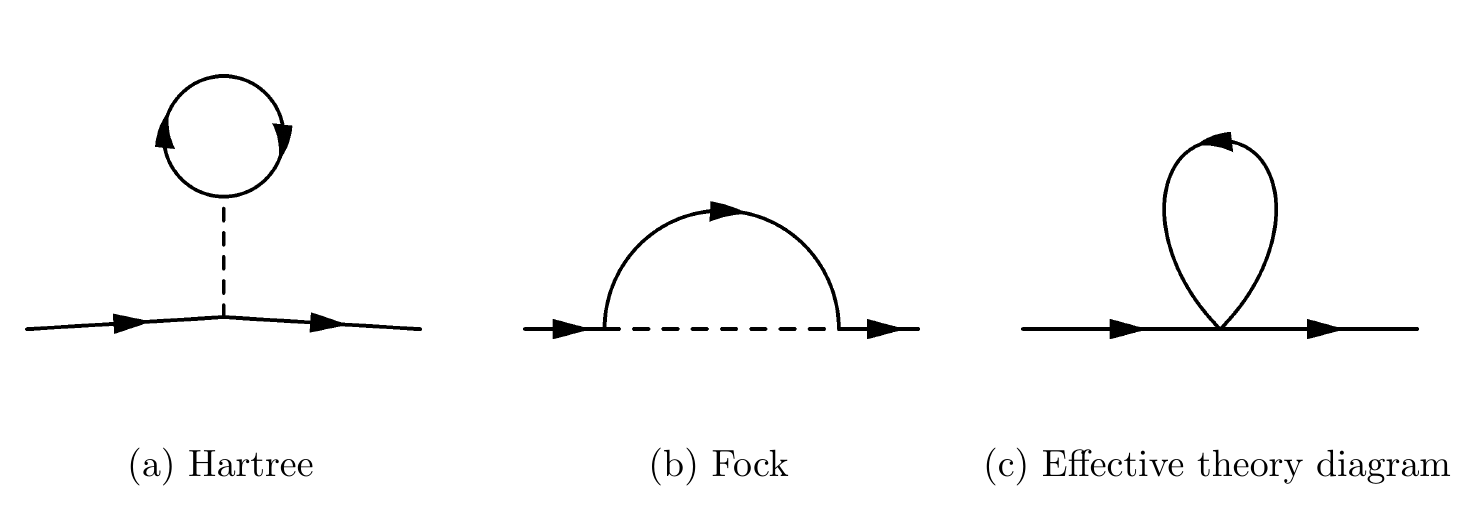}
\caption{The one-loop diagrams 
graphically representing the Hartree and Fock terms in the equation of motion~(\ref{2PI EoM: 1 loop}). 
These terms are obtained by 
variation of the effective action~(\ref{eq2PI.7}--\ref{2PI action: 2 loop}), 
whose diagrammatic representation is given in Figure~\ref{fig:fig2}.  
The dashed line corresponds to the torsion field, which reduces to a point in the effective theory.}
\label{fig:fig1}
\end{figure}
Consider now the one-loop corrected equation of motion for the fermionic two-point function, 
which is obtained by varying the action~(\ref{Action and counterterms}),
\begin{eqnarray}
\left (i\cancel{D} \!-\! m_R-im_I\gamma^5\right ) (iS^{ab}(x;x')) 
   &=&\sigma_z^{ab} i  \frac{\delta^D(x\!-\!x')}{\sqrt{-g}}
\label{2PI EoM: 1 loop}\\
&&\hskip -6.5cm
- 3 \pi G_N \xi^2 \Big\{
   {\rm Tr}\big[(iS^{aa}(x;x))\gamma^5\gamma^\sigma\big]
           \gamma^5\gamma_\sigma
   \!-\!\gamma^5\gamma^\sigma(iS^{aa}(x;x))\gamma^5\gamma_\sigma\Big\}(iS^{ab}(x;x'))
 \nonumber\\
&&\hskip -6.5cm + \big (\Delta\beta R + \Delta m_R \big )(iS^{ab}(x;x'))
\,,
\nonumber
\end{eqnarray}
where there is no summation over $a=\pm$.
The one-loop terms $\propto G_N \xi^2$ correspond to the Hartree and Fock contributions, respectively,
 to the one-loop effective equation
of motion, which are diagramatically illustrated in figure~\ref{fig:fig1}.

These contributions are local (in physical space), 
and we shall now show that a non-perturbative renormalization scheme -- akin to
a perturbative renormalization -- applies to the one-loop equation of motion~(\ref{2PI EoM: 1 loop}) for the full propagator
$S$. Since at one-loop only the coincident propagator $S(x;x)$ contributes~(\ref{2PI EoM: 1 loop}), methods developed 
for renormalization of the tree-level coincident propagator~\cite{Bunch, Christensen} apply.
For completeness, we note that the (free) fermionic propagator $S_0(x;x')$ obeys the following equation,
\begin{equation}
\label{2PI EoM: tree level}
\left (i\cancel{D} \!-\! m_R-im_I\gamma^5\right ) (iS_0(x;x')) 
   =\sigma_z i  \frac{\delta^D(x\!-\!x')}{\sqrt{-g}}
\,.
\end{equation}

The first step of the renormalization of~(\ref{2PI EoM: 1 loop}) 
is an UV expansion of the Schwinger-Keldysh propagator $S(x;x')$.  
Following the steps described in~\cite{Bunch, Christensen} we write, 
\begin{equation}
\label{eq2PI.3}
 S(x;x') = \left (i\cancel{D} \!+\! m_R\!-\!im_I \gamma^5\right ) \mathcal{G}(x;x') 
\,,
\end{equation}
where now $\mathcal{G}(x;x')$ is a Lorentz scalar spinor,
{\it i.e.} a spinor that depends on $\gamma^5$ and $\Sigma^{\mu\nu}=(1/4)[\gamma^\mu,\gamma^\nu]$
and a Lorentz scalar that depends on $\|x-x'\|^2$ only, the latter being justified for short distances. 
We are not sure whether the {\it Ansatz}~(\ref{eq2PI.3}) is correct 
for non-perturbative renormalization of the 2PI effective action~(\ref{eq2PI.7}--\ref{2PI action: 2 loop}).
However for the purpose of the perturbative renormalization that we are elaborating on here 
it is easy to convince oneself that the form~(\ref{eq2PI.3}) is valid not just for the free propagator 
$S_0(x;x')$ but also for the perturbatively corrected $S(x;x')$.
Note that the {\it Ansatz}~(\ref{eq2PI.3}) is only valid for short distances and does not work, in general, in the infrared. 
For example, in cosmological spaces time translation invariance is broken, and consequently 
one ought to treat separately the positive and negative frequency parts of the fermion propagator, see {\it e.g.} Ref.~\cite{Prokopec3}.
In practice that means that, in addition to the Lorentz invariant spinors, $\mathcal{G}(x;x')$ will contain 
positive and negative frequency projectors, $(1\pm\gamma^0)/2$. Of course, in more complicated 
space-times (such as black hole space-times) where more symmetries are broken,
 $\mathcal{G}(x;x')$ is expected to acquire in the infrared an even more complex Lorentz breaking spinorial structure.

 Nevertheless, in order to calculate the form of UV divergences of this theory, the {\it Ansatz} (\ref{eq2PI.3}) suffices.
Whatever is the actual contribution to the propagator in the infrared, we can always absorb it in the regular part of the propagator 
that we leave undetermined. Schematically, this amounts to breaking up the propagator as follows,
\begin{equation}
\begin{split}
S(x;x') =& S^{(\textrm{div})}(x;x') + S^{(\textrm{reg})}(x;x') 
   = \left (i\cancel{D}  \!+\! m_R\!-\! im_I \gamma^5 \right ) \mathcal{G}(x;x') \\
&+ \big (S^{(\textrm{div})}(x;x') - \left (i\cancel{D} \!+\! m_R\!-\! im_I \gamma^5\right ) \mathcal{G}(x;x')\big ) 
 + S^{(\textrm{reg})}(x;x') \\
\equiv&  \left (i\cancel{D}  \!+\! m_R\!-\! im_I \gamma^5 \right ) \mathcal{G}(x;x') +\tilde S^{(\textrm{reg})}(x;x')\
\,.
\end{split}
\label{regular part of propagator}
\end{equation}
On general grounds we know that the divergent part of the propagator is generated by short scale quantum fluctuations
of the vacuum, for which the propagator has the structure given in Eq.~(\ref{eq2PI.3}), and therefore one can always choose 
a renormalization scheme such that at short distances, 
\begin{equation}
 S^{(\textrm{div})}(x;x') - \left (i\cancel{D}  \!+\! m_R\!-\! im_I \gamma^5  \right ) \mathcal{G}(x;x')
\;\rightarrow\; 0
\,,
\qquad {\rm when }\;\; \|x-x'\|\rightarrow 0
\,.
\label{short distance divergence}
\end{equation}
Next, when~(\ref{eq2PI.3}) is inserted into the one-loop propagator equation~(\ref{2PI EoM: 1 loop}), one gets,
\begin{eqnarray}
\left (g^{\mu\nu} D_\mu D_\nu+\frac{1}{4} R+ |m|^2\right ) \mathcal{G}(x;x')
+({\rm one\!-\!loop\; terms})
   &=&-\sigma_z  \frac{\delta^D(x\!-\!x')}{\sqrt{-g}}
\label{2PI EoM: 1 loop:b}
\,,
\nonumber
\end{eqnarray}
where we used the well known relation $\left [D_\mu, D_\nu\right ] = \frac{1}{8} \left [\gamma_\sigma,\gamma_\lambda\right ] {R^{\sigma\lambda}}_{\mu\nu}$, and $|m| = \sqrt{m_R^2 + m_I^2}$. The main idea in~\cite{Bunch, Christensen} is to expand $\mathcal{G}(x;x')$ along the geodesic path $\sigma$ leading from $x$ to $x'$~\footnote{The mathematical framework, including the study of bi-tensors, {\it i.e.}
 tensors defined at two space-times points, is well discussed in~\cite{Christensen1}.}. One can then expand the metric and the connection in terms of purely geometrical quantities, evaluated at $x$. The existence of such an expansion is guaranteed by the 
equivalence principle, as it guarantees the existence of local metric expansions in terms of geometric quantities, 
a useful example being Riemann normal coordinates~\cite{Bunch}. 
Doing this procedure explicitly, we find the expansion and recursion relations~\cite{Christensen1},
\begin{subequations}
\begin{eqnarray}
\label{eq2PI.5.a} &&
\mathcal{G}(x;x') 
=\sum\limits_{n=0}^\infty\!\! - \Delta^{1/2}(x;x')\mathcal{A}_n (x;x') \int \frac{\text{d}^D k}{(2\pi)^D} 
e^{i k_\mu \sigma^{;\mu}}  \left (\!- \frac{\partial}{\partial |m|^2}\right )^n\!  \frac{1}{k^2\!-\!|m|^2 \!+\! i\epsilon} \,, 
\qquad\\
\label{eq2PI.5.b} &&\sigma_{;\rho} {{\mathcal{A}_{0}}_;}^\rho =0 \,, \\
\label{eq2PI.5.c} &&\sigma_{;\rho}  {{\mathcal{A}_{n+1}}_;}^\rho+ (n+1) \mathcal{A}_{n+1} = -\Delta^{-1/2} { \left ( \Delta^{1/2} \mathcal{A}_n\right )_{;\rho}}^\rho - \frac{1}{4} R \mathcal{A}_n\,,\quad
\end{eqnarray}
\end{subequations}
where the semicolon denotes covariant derivative, and $\Delta(x;x')$ is the Van Vleck-Morette determinant, given by\footnote{Note also that $\lim\limits_{x\rightarrow x'} \Delta(x;x') = 1$~\cite{Christensen1}.},
\[\Delta(x;x') = - g^{-1/2}(x) \text{det}( D_\mu D'_\nu \sigma(x;x')) g^{-1/2}(x')\,.\]
The object $\mathcal{A}_0$ is the spinor of parallel displacement. Its action on a spinor $\mathcal{A}(x;x')$ transports the spinor-at-$x'$ part of $\mathcal{A}(x;x')$ back to $x$. To calculate the coincidence limit of Eq. (\ref{eq2PI.5.a}), therefore, we have to form 
$\mathcal{A}_0 \mathcal{G}(x;x')$, a spinor-at-$x$, expanded in term of the geodesic distance $\sigma(x;x')$. Since we will need also derivatives, we want to construct $\mathcal{A}_0 {\mathcal{G}(x;x')}_{;\rho}$ too. Obviously $A_0^2 = \mathbb{1}$, and the expansion of $\mathcal{A}_0 {\mathcal{A}_0(x;x')}_{;\rho}$ is given in~\cite{Christensen}, and is of order $\mathcal{O}\left(\sigma(x;x')\right )$. 

The extra bit needed here is 
\begin{eqnarray}
\label{eq2PI.6.a}
 \mathcal{A}_0 \mathcal{A}_1 &=& -\frac{1}{12} R +\left [\frac{1}{24}  R_{;\rho} +\frac{1}{12} \Sigma_{[ab]} {{{R^{ab}}_{\rho\lambda}}_;}^\lambda\right ] {\sigma_;}^\rho + \mathcal{O}( \sigma^2)\,,\\
\label{eq2PI.6.b} \mathcal{A}_0 \mathcal{A}_{1;\mu} &=&- \frac{1}{24} R_{;\mu} +\frac{1}{12} \Sigma_{[ab]} {{{R^{ab}}_{\mu\lambda}}_;}^\lambda + \mathcal{O} (\sigma)\,,
\end{eqnarray}
where $\Sigma_{[ab]} = \frac{1}{4}  \left [\gamma_a, \gamma_b\right ]$. 
Taking coincidences limit is now trivial, we just need to send the geodesic distance $\sigma$~(and the geodesic tangent vector $\sigma_{;\mu}$) to zero.
We are now ready to calculate and regularize the vacuum divergences of this theory. We first show how to renormalize the equations of motion for the regular part of the propagator $\tilde S^{(\textrm{reg})}(x;x')$. 
Then we construct an effective 2PI action which regularizes the vacuum expectation value of the energy-momentum tensor.

By using the perturbative UV expansion of the full propagator that obeys~(\ref{2PI EoM: 1 loop}), 
by making use of the expansion~(\ref{eq2PI.5.a}--\ref{eq2PI.5.c}), 
we can calculate the first order divergent contribution to the coincident propagator. We get, 
\begin{eqnarray}
\left (i\cancel{D} \!-\! m_R-im_I\gamma^5\right ) (iS^{ab} (x;x')) 
   &=&\sigma_z^{ab} i  \frac{\delta^D(x\!-\!x')}{\sqrt{-g}}
\label{2PI EoM: 1 loop:b}\\
&&\hskip -6.7cm
+ 3 \pi G_N \xi^2 \bigg\{
\frac{\mu^{D-4}}{4\pi^2} \bigg [\frac{|m|^3}{2} \left (\frac{2}{D\!-\!4} \!+\! \gamma_E-1 \!+\! \log \frac{|m|^2}{4\pi\mu^2} \right ) 
      \!+\!\frac{|m| R(x)}{24}  \left (\frac{2}{D\!-\!4} \!+\! \gamma_E  \!+\! \log \frac{|m|^2}{4\pi\mu^2} \right ) \bigg ]
\nonumber  \\
&&\hskip -6.7cm
  - {\rm Tr}\big[(i\tilde S^{aa \rm (reg)}_0(x;x))\gamma^5\gamma^\sigma\big]
           \gamma^5\gamma_\sigma 
   \!+\!\gamma^5\gamma^\sigma(i\tilde S^{aa \rm (reg)}_0(x;x))\gamma^5\gamma_\sigma \bigg\}
                   (iS^{ab} (x;x'))
 \nonumber\\
&&\hskip -6.7cm  + \big (\Delta m_R+\Delta\beta R(x) \big )
            (iS^{ab}(x;x'))
\,,
\nonumber
\end{eqnarray}
where there is no summation over $a=\pm$.
By comparing the second line with the last line of this expression, we see that the following 
{\it minimal subtraction} choice, 
%
\begin{equation}
\label{Counter terms}  
\Delta m_R = -\frac{3G_N \xi^2 |m|^3}{4\pi} \frac{\mu^{D-4}}{D\!-\!4}  \,,\qquad
\Delta\beta = -\frac{G_N \xi^2|m|}{16\pi} \frac{\mu^{D-4}}{D\!-\!4}
\end{equation} 
subtracts the local divergences in~(\ref{2PI EoM: 1 loop:b}) resulting in,
\begin{eqnarray}
\left (i\cancel{D} \!-\! m_R-im_I\gamma^5\right ) (iS^{ab}(x;x')) 
   &=&\sigma_z^{ab} i  \frac{\delta^D(x\!-\!x')}{\sqrt{-g}}
\label{2PI EoM: 1 loop:c}\\
&&\hskip -6.5cm
+ 3 \pi G_N \xi^2 \bigg\{
\frac{1}{4\pi^2} \bigg [\frac{|m|^3}{2} \left ( \gamma_E-1 \!+\! \log \frac{|m|^2}{4\pi\mu^2} \right ) 
      \!+\!\frac{|m| R(x)}{24}  \left ( \gamma_E  \!+\! \log \frac{|m|^2}{4\pi\mu^2} \right ) \bigg ]
\nonumber  \\
&&\hskip -6.5cm
  - {\rm Tr}\big[(i\tilde S^{aa\rm (reg)}_0(x;x))\gamma^5\gamma^\sigma\big]
           \gamma^5\gamma_\sigma
   \!+\!\gamma^5\gamma^\sigma(i\tilde S^{aa\rm (reg)}_0(x;x))\gamma^5\gamma_\sigma\bigg\}
                     (iS^{ab}(x;x'))
\,.
\nonumber
\end{eqnarray}
Since all loop terms in this equation are regular, the divergences of the full propagator $S^{ab}(x;x')$ in (\ref{2PI EoM: 1 loop:c})
must be identical to those of the free propagator, which implies that the regular part of the dressed propagator obeys 
the following equation, 
\begin{eqnarray}
\left (i\cancel{D} \!-\! m_R-im_I\gamma^5\right ) (iS^{ab\rm (reg)}(x;x')) |_{x'\rightarrow x}
   &=&
\label{2PI EoM: 1 loop:d}\\
&&\hskip -7.7cm
+ 3 \pi G_N \xi^2 \bigg\{
\frac{1}{4\pi^2} \bigg [\frac{|m|^3}{2} \left ( \gamma_E-1 \!+\! \log \frac{|m|^2}{4\pi\mu^2} \right ) 
      \!+\!\frac{|m| R(x)}{24}  \left ( \gamma_E  \!+\! \log \frac{|m|^2}{4\pi\mu^2} \right ) \bigg ]
\nonumber  \\
&&\hskip -7.7cm
  - {\rm Tr}\big[(i\tilde S^{aa\rm (reg)}_0(x;x))\gamma^5\gamma^\sigma\big]
           \gamma^5\gamma_\sigma
   \!+\!\gamma^5\gamma^\sigma(i\tilde S^{aa\rm (reg)}_0(x;x))\gamma^5\gamma_\sigma\bigg\}
                     (iS^{ab\rm (reg)}(x;x))
.
\nonumber
\end{eqnarray}
This means that the regularization of the one-loop 
equation of motion for the full propagator is identical to the perturbative regularization of the same equation,
completing the perturbative renormalization of the one-loop 2PI equation of motion for the Keldysh propagator.

\begin{figure}
\centering
    \includegraphics[width=4in]{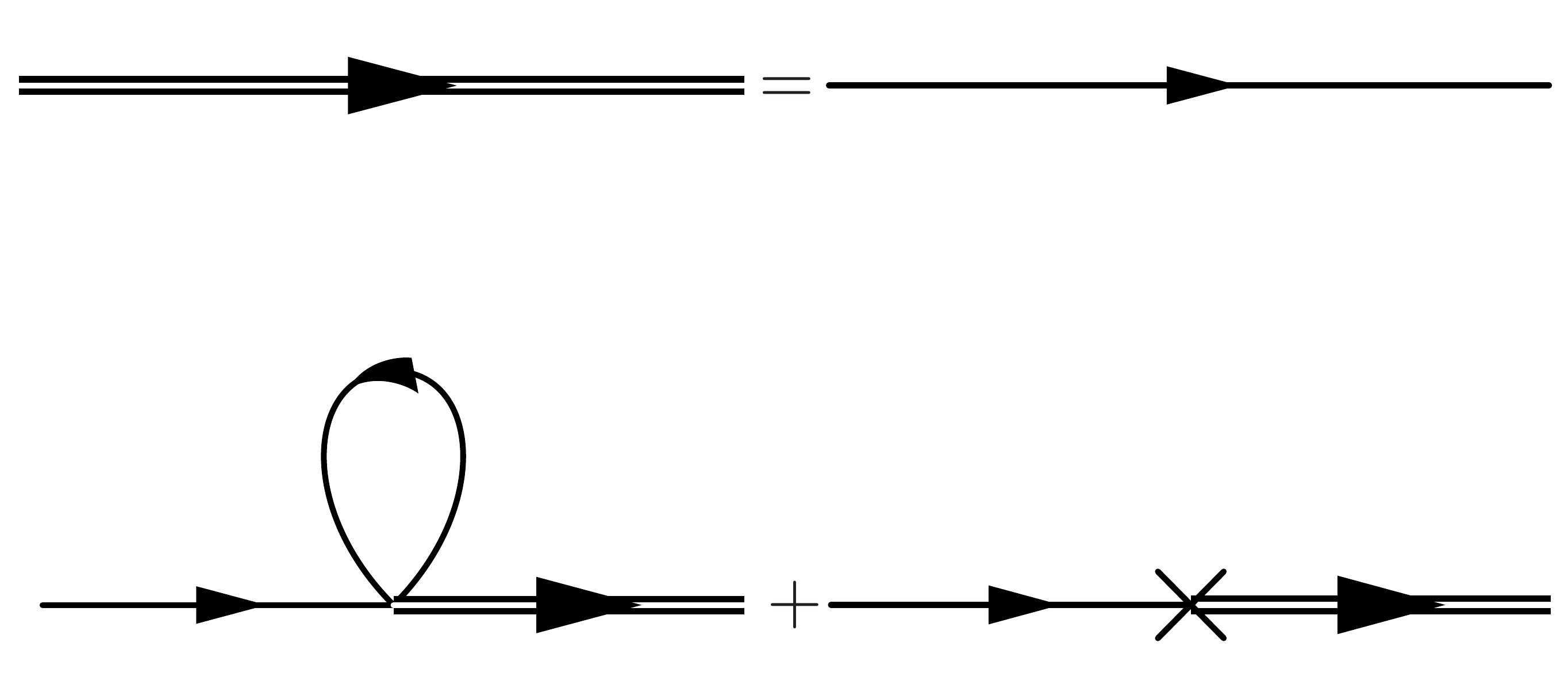}
\caption{The diagrammatic representation of the Dyson equation we renormalize in this paper. Note that the propagator in the loop is not the full one.
In fact, expanding the full propagator in a perturbative series as follows from this Dyson equation, it is possible to show 
that the counterterms we added in~(\ref{Action and counterterms}) takes care of all the divergences in this equation.}
\label{fig:Dyson.Eq}
\end{figure}
\begin{figure}
\centering
    \includegraphics[width=4in]{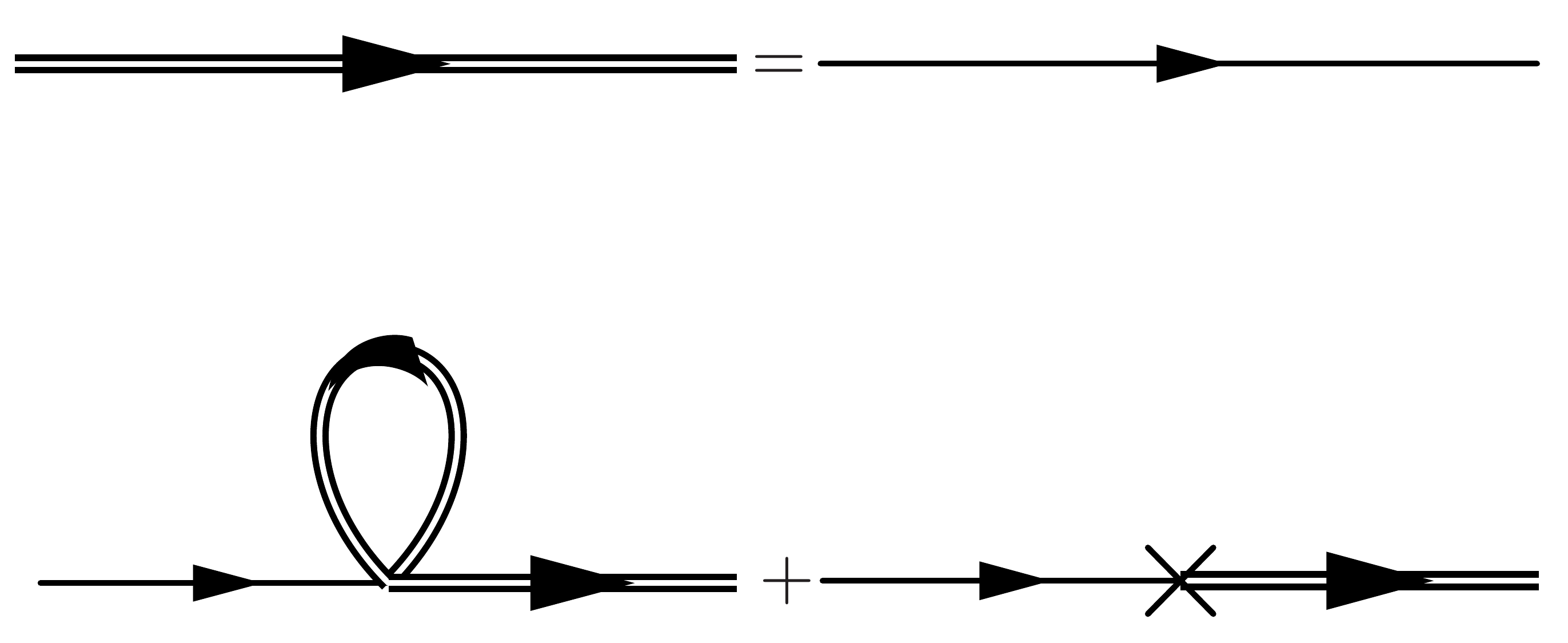}
\caption{The diagrammatic representation of the full Dyson equation. This equation needs to be renormalized non-perturbatively in fact, 
as it generates infinitely many divergent inequivalent diagrams.}
\label{fig:Dyson.Eq.Full}
\end{figure}

The procedure described here does not fully renormalize the equation of motion that we use in the main text, 
namely~(\ref{2PI EoM: 1 loop:d}). In Figure~\ref{fig:Dyson.Eq} we show the diagrammatic representation of the 
equation that we have renormalized. In the main text, we actually solve the full one-loop Dyson equation, 
which could be represented as in Figure~\ref{fig:Dyson.Eq.Full}. The divergencies that the loop expansion 
generated by Figure~\ref{fig:Dyson.Eq.Full} yields, cannot be cancelled by adding only one counter-term to the action
but require infinitely many insertions to the 2PI effective action~(\ref{Action and counterterms}). 
However, as we do with the first order correction, we can choose the renormalized parameters
to be small in Planck units. This means that we can disregard their contributions 
until the energy scale of $\mathcal{O}\left ( \frac{M^2_P}{\xi^{2/3}}\right )$, when our semi-classical approximation breaks down anyway.
We expect that all the other counter-terms, in the perturbative expansion, to be even more suppressed up to energy scales of 
$\mathcal{O}\left ( \frac{M^2_P}{\xi^{2/3}}\right )$. In principle these higher order corrections could 
be calculated: they would generate a series which should converge in the range in which the 
perturbative expansion is valid. This means that we can qualitatively trust our results, 
if the energy scale reached during the bounce is $< \frac{M^2_P}{\xi^{2}}$.

Note that the terms in the second line contribute as (time dependent) masses to the fermion,
while the terms in the third line are also local, but they have a more complicated structure.
The first (Hartree) term contributes as the axial vector field, while the interpretation of the latter (Fock) term 
depends on the spinorial structure of $(i\tilde S^{aa\rm (reg)}(x;x))$ and a detailed analysis 
shows that in a cosmological background the Fock term can give a contribution
to a (speudo-scalar) mass term, to a (pseudo-)vector current or to a spin density
(that couples to $\Sigma_{\mu\nu}$).

We now turn our attention to the energy momentum tensor and on how the counter-terms added in the 
2PI effective action~(\ref{Action and counterterms}) renormalize the Einstein's equations, 
as well as the propagator equation~(\ref{2PI EoM: 1 loop}). 
Varying the action~(\ref{Action and counterterms}) leads to the energy momentum tensor according to the definition,
\begin{eqnarray}
\label{non renormalized energy momentum tensor 2PI}
\langle T^{a}_{\mu\nu}\rangle &\equiv& \frac{2}{\sqrt{-g}} \frac{\delta }{\delta g_a^{\mu\nu}} \left ( \Gamma[S, g_{\mu\nu}] +S^{(\text{ct})}[ S, g_{\mu\nu}]\right )  \\
 &=&\frac{2}{\sqrt{-g}} \frac{\delta }{\delta g_a^{\mu\nu}}\left[\int \text{d}^Dx \sqrt{-g}\sigma_z \left ( \frac{\Delta\Lambda}{16\pi G_N}\! -\! \frac{\Delta G_N^{-1}}{16\pi } R\! +\! \Delta\alpha R^2\!+\! \Delta\zeta R_{\mu\nu}R^{\mu\nu}\right )\right]  \nonumber\\
 &&\!-\frac{2}{\sqrt{-g}} \frac{\delta }{\delta g_a^{\mu\nu}}\left[\int \text{d}^Dx \sqrt{-g} \sigma_z \left (\Delta m_R +\Delta\beta R\right )\right] (iS(x;x)[g])\nonumber \\
 &&\!-\frac{3\pi G\xi^2}{2}g^a_{\mu\nu}  \sigma_z^{aa}\Big\{
   {\rm Tr}\big[(iS^{aa}(x;x)[g])\gamma^5\gamma_\sigma\big]{\rm Tr}\big[\gamma^5\gamma^\sigma(iS^{aa}(x;x)[g])\big]
           \nonumber\\
   &&\!-{\rm Tr}\big[\gamma^5\gamma^\sigma(iS^{aa}(x;x)[g])\gamma^5\gamma_\sigma(iS^{aa}(x;x)[g])\big]\Big\}\nonumber\\
   &&\!+\sigma_z \frac{2}{\sqrt{-g}} \left[\frac{\delta }{\delta g_a^{\mu\nu}} \int \text{d}^Dx \sqrt{-g} \left(i\cancel{D} - m_R - i\gamma^5 m_I\right )\right](iS^{aa}(x;x)[g])\nonumber
   \,,
\end{eqnarray}
where $(iS(x;x)[g])$ satisfies Eq.~(\ref{2PI EoM: 1 loop}) and there is no summation over $a=\pm$. 
Note that the energy-momentum tensor acquires a polarities structure, depending on whether the metric field
that couples to it lies on the upper or lower branch of the contour from figure~\ref{fig:Image2PIbis}.
Eventually, since $g^a_{\mu\nu}$ is a classical field, and since the fermionic propagators in the energy momentum tensor are 
evaluated at coincidence, this distinction becomes irrelevant to solve the semi-classical Einstein's equations,
as the two equations are not linearly independent. 
We decide nevertheless to make it explicit in here, to explain why the geometrical counter-terms in the 2PI effective action~(\ref{Action and counterterms})
appear multiplied by $\sigma_z$.

Note further that the energy-momentum tensor is an {\it on-shell} quantity, such that $(iS(x;x)[g])$ is actually a functional of the metric,
as follows from Eq.~(\ref{2PI EoM: 1 loop}). We have made this explicit by writing $(iS(x;x)[g])$, instead of $(iS(x;x))$. 
Now we go back to Eq.~(\ref{2PI EoM: 1 loop}): multiplying each side by $\sigma_z\sqrt{-g}$, integrating with respect to $x$ and taking the variation with respect to $g^a_{\mu\nu}$
we find,
\begin{eqnarray}
\label{integral propagator equation}
\left[\frac{\delta }{\delta g_a^{\mu\nu}} \int\text{d}^D x \sqrt{-g} \sigma_z\left (i\cancel{D} \!-\! m_R-im_I\gamma^5\right )\right] (iS^{ab}(x;x')[g]) 
   &=& \\
   &&\hskip -10cm- \int\text{d}^D x \sqrt{-g} \sigma_z\left (i\cancel{D} \!-\! m_R-im_I\gamma^5\right )\left[\frac{\delta }{\delta g_a^{\mu\nu}}(iS^{ab}(x;x')[g]) \right]\nonumber
\\
&&\hskip -10cm
- 3 \pi G_N \xi^2\frac{\delta }{\delta g_a^{\mu\nu}} \Bigg[\int\text{d}^D x  \sqrt{-g} \sigma_z\Big\{
   {\rm Tr}\big[(iS^{aa}(x;x)[g])\gamma^5\gamma^\sigma\big]
           \gamma^5\gamma_\sigma\nonumber\\
  &&\hskip -10cm \!-\!\gamma^5\gamma^\sigma(iS^{aa}(x;x)[g])\gamma^5\gamma_\sigma\Big\} 
  (iS^{ab}(x;x')[g])\nonumber\\
  &&\hskip -10cm \!+\! \int\text{d}^D x  \sqrt{-g} \sigma_z\big (\Delta\beta R + \Delta m_R \big )(iS^{ab}(x;x')[g])\Bigg] \nonumber
  \,,
\end{eqnarray}
which we can plug back into the last line energy-momentum tensor~(\ref{non renormalized energy momentum tensor 2PI}), 
to simplify the variation of the tree-level term. We then obtain 
\begin{eqnarray}
\label{2PI effective action, onshell expression}
\langle T^a_{\mu\nu}\rangle &=& \frac{2}{\sqrt{-g}}\Bigg\{ \frac{\delta }{\delta g_a^{\mu\nu}}\bigg[\int \text{d}^Dx \sqrt{-g}\sigma_z \left ( \frac{\Delta\Lambda}{16\pi G_N}\! -\! \frac{\Delta G_N^{-1}}{16\pi } R\! +\! \Delta\alpha R^2\!+\! \Delta\zeta R_{\mu\nu}R^{\mu\nu}\right )\nonumber\\
&&-  \frac{3 \pi G_N \xi^2}{2} \sqrt{-g}\sum\limits_{a,b=\pm}\sigma^{ab}_z\Big\{
   {\rm Tr}\big[(iS^{aa}(x;x)[g])\gamma^5\gamma^\sigma\big]
           {\rm Tr}\big[\gamma^5\gamma_\sigma(iS^{aa}(x;x)[g])\big]\nonumber\\
  &&\!-\!{\rm Tr}\big[\gamma^5\gamma^\sigma(iS^{aa}(x;x)[g])\gamma^5\gamma_\sigma
  (iS^{aa}(x;x)[g])\big]\Big\} \bigg]\nonumber\\
  &&- {\rm Tr}\Bigg[\int\text{d}^D x \sqrt{-g} \sum\limits_{a,b=\pm}\sigma^{ab}_z\Bigg(i\cancel{D} \!-\! m_R-im_I\gamma^5-\Delta m_R -\Delta\beta R \nonumber\\
  &&+ 3\pi G_N \xi^2\Big\{
   {\rm Tr}\big[(iS^{aa}(x;x)[g])\gamma^5\gamma^\sigma\big]
           \gamma^5\gamma_\sigma\nonumber\\
  && \label{2PI effective action, onshell expression:a}\!-\!\gamma^5\gamma^\sigma(iS^{aa}(x;x)[g])\gamma^5\gamma_\sigma\Bigg)\frac{\delta }{\delta g_a^{\mu\nu}}(iS^{ab}(x;x)[g])\Bigg ]\Bigg\}\\
 &=& \frac{2}{\sqrt{-g}}\frac{\delta }{\delta g_a^{\mu\nu}} \Bigg\{ \int \text{d}^Dx \sqrt{-g}\sigma_z\bigg(\sigma_z\frac{\Delta\Lambda}{16\pi G_N}\! -\! \sigma_z\frac{\Delta G_N^{-1}}{16\pi } R\! +\! \sigma_z\Delta\alpha R^2\nonumber \\
  &&- \frac{3 \pi G_N \xi^2}{2} \sum\limits_{a,b=\pm}\sigma^{ab}_z\Big\{
   {\rm Tr}\big[(iS^{aa}(x;x)[g])\gamma^5\gamma^\sigma\big]
          {\rm Tr}\big[\gamma^5\gamma_\sigma(iS^{aa}(x;x)[g])\big]\nonumber\\
  &&\!-\!{\rm Tr}\big[\gamma^5\gamma^\sigma(iS^{aa}(x;x)[g])\gamma^5\gamma_\sigma
  (iS^{aa}(x;x)[g])\big]\Big\} \nonumber\\ 
  &&-\frac{i}{\sqrt{-g}} {\rm Tr}\log (iS(x;x)[g])\bigg)\Bigg\}\nonumber\\
  \label{2PI effective action, onshell expression, definiton}&\equiv& \frac{2}{\sqrt{-g}} \frac{\delta }{\delta g_a^{\mu\nu}} \Gamma_{\rm Eff}[(iS[g]), g]
  \,,
\end{eqnarray}
where we used the relations,
\begin{eqnarray}
\label{Inverse propagator definition on shell}
\int\text{d}^Dx'' (iS^{-1,ac}(x,x''))(iS^{cb}(x'',x'))&=& \delta^D(x-x')\delta^{ab}\,, \\
\int\text{d}^Dx'' (iS^{-1}(x,x'')[g])\frac{\delta}{\delta g^{\mu\nu}}(iS(x'',x')[g]) &=& \frac{\delta}{\delta g^{\mu\nu}} \log (iS(x,x')[g] )\,, \nonumber\\
(iS^{-1, ab}(x',x)[g]) =-i\sigma^{ab}_z\sqrt{-g(x)}\Bigg(i\cancel{D} \!-\! m_R-im_I\gamma^5&&\hskip-0.7cm-\Delta m_R -\Delta\beta R \nonumber\\
  &&\hskip -5cm+ 3\pi G_N \xi^2\Big\{
   {\rm Tr}\big[(iS^{aa}(x;x)[g])\gamma^5\gamma^\sigma\big]
           \gamma^5\gamma_\sigma\nonumber\\
  &&\hskip -5cm\!-\!\gamma^5\gamma^\sigma(iS^{aa}(x;x)[g])\gamma^5\gamma_\sigma\Bigg)\delta^D(x'-x) \,.\nonumber
\end{eqnarray}
%
The proof provided by~(\ref{2PI effective action, onshell expression}--\ref{2PI effective action, onshell expression, definiton}) 
constitutes a formal derivation of the two-loop effective action, evaluated on shell, whose variation with respect to the metric
gives $\langle T^a_{\mu\nu}\rangle$. It is in fact worth noting that the action defined in Eq.~(\ref{2PI effective action, onshell expression:a}) is 
actually nothing different than the action we started with, (\ref{Action and counterterms}), evaluated {\it on-shell}. 
However, this is not yet clear from Eq.~(\ref{2PI effective action, onshell expression:a}), but it will become more apparent 
once the perturbative solution to Eq.~(\ref{2PI EoM: 1 loop}) is used to evaluate the vacuum divergencies of the energy momentum tensor in~(\ref{2PI effective action, onshell expression}).

Renormalizing the effective action~(\ref{2PI effective action, onshell expression}) can now be done 
{\it on-shell}, since we proved that the variation with respect to the metric of the effective action defined in Eq.~(\ref{2PI effective action, onshell expression})
gives the correct {\it on-shell} expression for $\langle T^a_{\mu\nu}\rangle$.
Note that the one-loop term, {\it i.e.} ${\rm Tr}\log (iS(x;x)[g])$, is calculated with respect to 
the full propagator, not just the free one, and the strangely opposite sign, 
when compared to~(\ref{Action and counterterms}), of the two-loop term $\propto \frac{3 \pi G_N \xi^2}{2}$.
These two details are actually of critical importance for renormalization, as we shall see shortly. 

Consider the perturbative solution of Eq.~(\ref{2PI EoM: 1 loop}), at first order in $ 3 \pi G_N \xi^2$,
\begin{eqnarray}
\label{perturbative expansion 1}  &&(iS^{ab}(x;x')[g]) \!=\!   (iS^{ab}_0(x;x')[g]) \qquad\nonumber\\
     && \!+\! 3 \pi G_N \xi^2i \int\text{d}^D x'' \sqrt{-g(x'')}(iS^{ac}_0(x;x'')[g])\sigma_z^{cd} \Big\{ {\rm Tr}\big[(iS_0^{dd}(x'';x'')[g])\gamma^5\gamma^\sigma\big]
           \gamma^5\gamma_\sigma\nonumber \\
&&\hskip 5.1cm\!-\gamma^5\gamma^\sigma (iS_0^{dd}(x'';x'')[g])\gamma^5\gamma_\sigma\Big\}
     (iS^{db}_0(x'';x')[g])\nonumber \\
 &&-i\int\text{d}^D x''  \sqrt{-g(x'')}(iS^{ac}_0(x;x'')[g]) \sigma^{cd}_z\big (\Delta\beta R(x'') + \Delta m_R \big )(iS^{db}_0(x'';x')[g]) \nonumber\\
 &&+\mathcal{O}( 3 \pi G_N \xi^2)^2\,, \qquad
\end{eqnarray}
where $(iS^{ab}_0(x;x'))$ obeys the tree level equation~(\ref{2PI EoM: tree level}). Inserting the perturbative solution of 
Eq.~(\ref{2PI EoM: 1 loop}) into the effective action defined in~(\ref{2PI effective action, onshell expression}), and expanding the 
logarithm term as,
\begin{eqnarray*}
\log (iS^{\rm(div),ab}(x;x)[g]) &\simeq& \log (iS^{ab}_0(x;x)[g]) \!+\! 3 \pi G_N \xi^2i  \!\sqrt{-g(x)}\sigma_z^{ad} \nonumber\\
&&\hskip -3.5cm\times\Big\{ {\rm Tr}\big[(iS_0^{dd}(x;x)[g])\gamma^5\gamma^\sigma\big]
           \gamma^5\gamma_\sigma
\!-\! \gamma^5\gamma^\sigma (iS_0^{dd}(x;x)[g])\gamma^5\gamma_\sigma\Big\}
     (iS^{db}_0(x;x)[g])\nonumber \\
 &&\hskip -3.5cm\!-i  \sqrt{-g(x)} \sigma^{ad}_z\big (\Delta\beta R(x) + \Delta m_R \big )(iS^{db}_0(x;x)[g]) \,, 
\end{eqnarray*}
we finally arrive at the expression for the 2PI effective action, whose variation gives the {\it on-shell} energy momentum tensor: 
\begin{eqnarray}
\label{onshell 2PI effective action loop expansion}
 &&\hskip -3cm\Gamma_{\rm Eff}[(iS[g]), g]=\int \text{d}^Dx \sqrt{-g}\Bigg(\sigma_z\frac{\Delta\Lambda}{16\pi G_N}\! -\! \sigma_z\frac{\Delta G_N^{-1}}{16\pi } R\! +\! \sigma_z\Delta\alpha R^2 \!+\! \Delta\zeta R_{\mu\nu}R^{\mu\nu}\nonumber\\
  &&\hskip -3cm+ \frac{3 \pi G_N \xi^2}{2} \sum\limits_{a,b=\pm}\sigma^{ab}_z\Big\{
   {\rm Tr}\big[(iS_0^{aa}(x;x)[g])\gamma^5\gamma^\sigma\big]
          {\rm Tr}\big[\gamma^5\gamma_\sigma(iS_0^{aa}(x;x)[g])\big]\nonumber\\
  &&\hskip -3cm\!-\!{\rm Tr}\big[\gamma^5\gamma^\sigma(iS_0^{aa}(x;x)[g])\gamma^5\gamma_\sigma
  (iS_0^{aa}(x;x)[g])\big] \nonumber\\ 
  &&\hskip -3cm-\sigma^{ab}_z\big (\Delta\beta R(x) + \Delta m_R \big )(iS^{aa}_0(x;x)[g])\Big\}-\frac{i}{\sqrt{-g}} {\rm Tr}\log (iS_0(x;x)[g])\Bigg)
  \,.
\end{eqnarray}
Notice that inserting the perturbative expansion~(\ref{perturbative expansion 1}) into the action at the beginning in this 
appendix, Eq.~(\ref{Action and counterterms}), leads to the same result as~(\ref{onshell 2PI effective action loop expansion}), 
at linear order in $3\pi G_N \xi^2$. The two-loop contribution from the one-loop term, ${\rm Tr}\log (iS(x;x)[g])$,
and the two-loop contribution from the tree-level term, $(i\cancel{D} -m_R - i \gamma^5 m_I) (iS_0(x;x)[g])$,
in fact come with the opposite sign, and cancel each other. The only two-loop divergent part of the effective action~(\ref{Action and counterterms}) 
is given by the (sum of all) 2PI irreducible diagrams with one vertex and two propagators, pictured in Figure~\ref{fig:fig2}, 
and it is given by evaluating $\Gamma_2[(iS_0(x;x)[g]), g]$ as in~(\ref{eq1.15}). 

Renormalizing the one-loop contribution, {\it i.e.} ${\rm Tr}\log (iS_0(x;x)[g])$, would yield to the addition to the effective action~(\ref{onshell 2PI effective action loop expansion})
of the terms~\cite{Christensen},
\begin{equation}
\label{one-loop contribution} \Gamma^{\rm (one-loop)}_{\rm (ren)}\! = \!\int\text{d}^4 x \sqrt{-g} \bigg (\frac{\Lambda^{\rm (1), (ren)}}{16\pi G_N} -\frac{R}{16\pi G^{\rm (1), (ren)}_N} +\alpha^{\rm (1), (ren)} R^2 +\zeta^{\rm (1), (ren)} R_{\mu\nu}R^{\mu\nu}\bigg) \,.
\end{equation}
Here the renormalized coupling constants, $\Lambda^{\rm (1), (ren)}, \,G^{\rm (1), (ren)}_N\,,\alpha^{\rm (1), (ren)}\,,\zeta^{\rm (1), (ren)}$, 
need to be determined by experiments and they possess the same scale dependence we will find for the two-loop 
renormalized parameters, {\it i.e.} $\log \frac{|m|^2}{4\pi\mu^2}$. 

In the effective action~(\ref{onshell 2PI effective action loop expansion}), we want to perform the familiar splitting 
defined in Eq.~(\ref{regular part of propagator}). This would lead to,
\begin{eqnarray}
\label{onshell 2PI effective action loop expansion: div reg split}
&&\hskip -1cm \Gamma_{\rm Eff}[(iS^{\rm (div)}[g]),(iS^{\rm (reg)}[g]), g]=\int \text{d}^Dx \sqrt{-g}\Bigg(\sigma_z\frac{\Delta\Lambda}{16\pi G_N}\! -\! \sigma_z\frac{\Delta G_N^{-1}}{16\pi } R\! +\! \sigma_z\Delta\alpha R^2 \nonumber\\
  &&\hskip -1cm+ \frac{3 \pi G_N \xi^2}{2} \!\!\sum\limits_{a,b=\pm}\sigma^{ab}_z\Big\{
   {\rm Tr}\big[(iS_0^{aa,\rm (reg)}(x;x)[g])\gamma^5\gamma^\sigma\big]
          {\rm Tr}\big[\gamma^5\gamma_\sigma(iS_0^{aa\rm (reg)}(x;x)[g])\big]\nonumber\\
  &&\hskip -1cm\!-\!{\rm Tr}\big[\gamma^5\gamma^\sigma(iS_0^{aa,\rm (div)}(x;x)[g])\gamma^5\gamma_\sigma
  (iS_0^{aa,\rm (div)}(x;x)[g])\big] \nonumber\\ 
   &&\hskip -1cm\!-\!2{\rm Tr}\big[\gamma^5\gamma^\sigma(iS_0^{aa,\rm (div)}(x;x)[g])\gamma^5\gamma_\sigma
  (iS_0^{aa,\rm (reg)}(x;x)[g])\big] \nonumber\\ 
   &&\hskip -1cm\!-\!{\rm Tr}\big[\gamma^5\gamma^\sigma(iS_0^{aa,\rm (reg)}(x;x)[g])\gamma^5\gamma_\sigma
  (iS_0^{aa,\rm (reg)}(x;x)[g])\big] \nonumber\\ 
  &&\hskip -1cm+\big (\Delta\beta R(x) + \Delta m_R \big )(iS^{\rm (div),aa}_0(x;x)[g]\!+iS^{\rm (reg),aa}_0(x;x)[g])\Big\}\Bigg) \\
  &&\hskip -1cm+\Gamma^{\rm (one-loop)}_{\rm (ren)} \nonumber
  \,.
\end{eqnarray}
It is not hard to see that the term mixing the regular and divergent parts of the propagator appears,
and the regular part of the counter-terms $\propto \Delta\beta,\,\Delta m_R$, 
come with the same relative minus sign as in Eq.~(\ref{2PI EoM: 1 loop}). 
Noting the extra factor of 2 in the term mixing the regular and divergent part of the propagator, 
makes this particular divergence finite, in the~{\it minimal subtraction} prescription,
by the same choice of $\Delta m_R$ and $\Delta\beta$ given in~(\ref{Counter terms}).

We are therefore only left with the divergent contribution,
\begin{eqnarray}
\label{onshell 2PI effective action loop expansion: div reg split: a} 
\int \text{d}^Dx \sqrt{-g}\Bigg(\sigma_z\frac{\Delta\Lambda}{16\pi G_N}\! -\!\sigma_z \frac{\Delta G_N^{-1}}{16\pi } R\! +\!\sigma_z \Delta\alpha R^2 \nonumber\\
 &&\hskip -7cm- \frac{3 \pi G_N \xi^2}{2} \!\!\sum\limits_{a,b=\pm}\sigma^{ab}_z\Big\{{\rm Tr}\big[\gamma^5\gamma^\sigma(iS_0^{aa,\rm (div)}(x;x)[g])\gamma^5\gamma_\sigma
  (iS_0^{aa,\rm (div)}(x;x)[g])\big] \nonumber\\
  &&\hskip -7cm+\big (\Delta\beta R(x) + \Delta m_R \big )(iS^{\rm (div),aa}_0(x;x)[g])\Big\}\Bigg)
   \,,
\end{eqnarray}
which can be rewritten using the relations
\begin{equation}
\begin{split}
\label{square completion relations}
 \frac{\Delta\Lambda}{16\pi G_N} =&- \frac{(\Delta m_R)^2}{6\pi G_N \xi^2} \,,\qquad
\Delta\alpha =- \frac{(\Delta\beta)^2}{6\pi G_N \xi^2} \,,\qquad
\frac{\Delta G_N^{-1}}{16\pi } = \frac{\Delta m_R\Delta\beta}{3\pi G_N \xi^2} \,,
\end{split}
\end{equation}
as
\begin{eqnarray}
 -\int \text{d}^Dx \sqrt{-g} \sigma_z\Bigg\{\frac{\Delta m_R^2}{6\pi G_N} +\frac{\Delta\beta^2}{6\pi G_N}R^2 +\frac{2\Delta m_R\Delta\beta}{6\pi G_N} \nonumber \\
 &&\hskip -7cm-(\Delta\beta R +\Delta m_R)\bigg[\frac{\mu^{D-4}}{D-4}\bigg(\frac{|m|^3}{2} \left (\frac{2}{D\!-\!4} \!+\! \gamma_E-1 \!+\! \log \frac{|m|^2}{4\pi\mu^2} \right ) \nonumber\\
 &&\hskip -7cm+\frac{|m| R(x)}{24}  \left (\frac{2}{D\!-\!4} \!+\! \gamma_E  \!+\! \log \frac{|m|^2}{4\pi\mu^2} \bigg ) \right)\bigg] \nonumber\\
 &&\hskip -7cm+\frac{3 \pi G_N \xi^2}{2}\bigg[\frac{\mu^{D-4}}{D-4}\bigg(\frac{|m|^3}{2} \left (\frac{2}{D\!-\!4} \!+\! \gamma_E-1 \!+\! \log \frac{|m|^2}{4\pi\mu^2} \right )\nonumber\\
 &&\hskip -7cm+\frac{|m| R(x)}{24}  \left (\frac{2}{D\!-\!4} \!+\! \gamma_E  \!+\! \log \frac{|m|^2}{4\pi\mu^2} \bigg ) \right)\bigg]^2\Bigg\} \nonumber\\
 &&\hskip -7cm= -\int \text{d}^Dx \frac{\sqrt{-g}}{6\pi G_N \xi^2}\sigma_z\Bigg (\Delta\beta R + \Delta m_R\nonumber\\
 &&\hskip -7cm - 3\pi G_N \xi^2\bigg[\frac{\mu^{D-4}}{D-4}\bigg(\frac{|m|^3}{2} \left (\frac{2}{D\!-\!4} \!+\! \gamma_E-1 \!+\! \log \frac{|m|^2}{4\pi\mu^2} \right )\nonumber\\
 \label{End renormalization}&&\hskip -7cm+\frac{|m| R(x)}{24}  \left (\frac{2}{D\!-\!4} \!+\! \gamma_E  \!+\! \log \frac{|m|^2}{4\pi\mu^2} \bigg ) \right)\bigg]\Bigg)^2
 \,,
 \end{eqnarray}
which is again renormalized by the choice given in~(\ref{Counter terms}), 
as the same combination, already encountered in Eq.~(\ref{2PI EoM: 1 loop}), appears in~(\ref{End renormalization}). 
This concludes the perturbative renormalization of the 2PI effective action~(\ref{Action and counterterms}).

To conclude this chapter, we report the renormalized 2PI effective action, as we will use it in the main text,
\begin{eqnarray}
\label{Action and counterterms: end}
&&\Gamma_{\rm ren}[S, g_{\mu\nu}] = \nonumber \\
&=& \int\!\text{d}^D x \sqrt{-g}\sigma_z \left [
   \frac{\Lambda^{\rm (ren)}}{16\pi G_N}\! -\! \frac{ G_N^{-1 \rm (ren)}}{16\pi } R\! +\! \alpha^{\rm (ren)} R^2 \!+\! \zeta^{\rm (1), (ren)} R_{\mu\nu}R^{\mu\nu}
       \!-\!\beta^{\rm (ren)} R \big(iS^{\rm (reg)}(x;x)\big) 
     \right ]
 \nonumber\\
      &&+\int\!\text{d}^D x \sqrt{-g}\!\Big[\sigma_z\text{Tr}[ \left (i\cancel{D}\! -\! m^{\rm (ren)}_R\! -\! im_I\gamma^5 \right ) (iS^{\rm (reg)}(x;x))]\Big] \nonumber\\
      &&+ \frac{3 \pi G_N \xi^2}{2}\sum_{a,b=\pm}\!\int\!\text{d}^D x \sqrt{-g}
   \Big[\sigma_z^{ab}{\rm Tr}\big[(iS^{bb, \rm (reg)}(x;x))\gamma^5\gamma^\sigma\big]
       {\rm Tr}\big[(iS^{bb, \rm(reg)}(x;x))\gamma^5\gamma_\sigma\big]
\nonumber\\
       &&
   -\sigma_z^{ab} {\rm Tr}\big[(iS^{bb, \rm(reg)}(x;x))\gamma^5\gamma^\sigma(iS^{bb, \rm(reg)}(x;x))\gamma^5\gamma_\sigma\big] + \mathcal{O}\left ( \xi^{4}/M^4_P\right ) \\
   &=&  \int\!\text{d}^D x \sqrt{-g}\sigma_z \left [
   \frac{\Lambda^{\rm (ren)}}{16\pi G_N}\! -\! \frac{ G_N^{-1 \rm (ren)}}{16\pi } R\! +\! \alpha^{\rm (ren)} R^2  \!+\! \zeta^{\rm (1), (ren)} R_{\mu\nu}R^{\mu\nu}
       \!-\! \beta^{\rm (ren)} R \big(iS^{\rm (reg)}(x;x)\big) 
     \right ]
 \nonumber\\
      &&+\int\!\text{d}^D x \sqrt{-g}\!\Big[\sigma_z\text{Tr}[ \left (i\cancel{D}\! -\! m^{\rm (ren)}_R\! -\! im_I\gamma^5 \right ) (iS^{\rm (reg)}(x;x))]\Big] \nonumber\\
      &&+\frac{3 \pi G_N \xi^2}{2}\int\!\text{d}^D x \sqrt{-g}{\rm Tr} \left\{
   \sigma_z i\Pi^{\rm (reg)}(x,x) iS^{\rm (reg)}(x;x) \right\}+\mathcal{O}\left ( \xi^{4}/M^4_P\right )\,,\nonumber
\qquad
\end{eqnarray}
 where the extra corrections of $\mathcal{O}\left ( \xi^{4}/M^4_P\right )$ denotes the rest of the counterterms needed to 
 renormalize the Dyson equation from Figure~\ref{fig:Dyson.Eq.Full}, and we absorbed the one loop contribution from Eq.~(\ref{one-loop contribution})
 into the renormalized parameters, $\Lambda^{\rm (ren)},\,G_N^{-1 \rm (ren)},\,\alpha^{\rm (ren)}$. 
 Note that the variation of this action with respect to the regular part of the propagator $\big(iS^{\rm(reg)}(x;x)\big)$
 correctly gives the renormalized equation (\ref{2PI EoM: 1 loop:d}). The {\it off-shell} variation of~(\ref{Action and counterterms: end}) with respect to the 
 metric, gives the energy momentum tensor, as reported in Eq.~(\ref{renormalized einstein's equations}). 
 
We conclude this appendix with a final remark. Notice that, in defining the {\it on-shell} effective action~(\ref{2PI effective action, onshell expression, definiton})
we have to invert the Dirac operator as in~(\ref{Inverse propagator definition on shell}). However, we can redefine
the propagator in~(\ref{Inverse propagator definition on shell}), as
 \begin{eqnarray}
 \label{inverse propagator redefinition}
&& \int\text{d}^Dx'' (iS^{-1,ac}(x,x''))(i\tilde{S}^{cb}(x'',x'))=\\
&=& \int\text{d}^Dx'' (iS^{-1,ac}(x,x''))(iS^{cb}(x'',x')-iS^{cb, \rm(reg)}(x,x)[g])= \delta^D(x-x')\delta^{ab}\,.
 \end{eqnarray}
 We would then find a different expression for $\tilde{\Gamma}_{\rm Eff}[(i\tilde{S}[g]), g]$, which would, however, still be 
 renormalized by the counterterms in~(\ref{Action and counterterms}). This is so because the redefinition~(\ref{inverse propagator redefinition}) is 
 purely a choice of different initial conditions for the solution to the propagator equation~(\ref{2PI EoM: 1 loop}).  In fact, since $(iS^{\rm(reg)}(x,x)[g])$ is a solution of~(\ref{2PI EoM: 1 loop:d}), 
 and because the Dirac operator in~(\ref{2PI EoM: 1 loop}) is linear, we 
 can always add a regular propagator to a solution of~(\ref{2PI EoM: 1 loop}) and still get a solution of~(\ref{2PI EoM: 1 loop:c}).
 It is quite obvious that renormalization results does not depend on the choice of initial conditions, 
 which is in fact what we find. It can be shown that $\tilde{\Gamma}_{\rm Eff}[(iS[g]), g]$ 
 is perturbatively renormalized by the same precedure described in this appendix. 
 Since $\tilde{\Gamma}_{\rm Eff}[(iS[g]), g]$ and $\Gamma_{\rm Eff}[(iS[g]), g]$
 are related by a change in the choice of initial conditions, and are both renormalized by the same counterterms,
 we can conclude that the form of the counterterms~(\ref{Counter terms}--\ref{square completion relations})
 is the same for every initial state.


\section{Mode functions renormalization} 
\label{Mode Functions}

In this section we show how to obtain the same result as the previous section, but instead of using the propagator expansion~(\ref{eq2PI.5.a}), we will solve the equations for the fermionic mode functions directly. This procedure is substantially different from what we have done in appendix~\ref{Propagators.Expansion}, because of the following. In appendix~\ref{Propagators.Expansion} we expanded the propagator and the effective action in powers of $\frac{1}{m^2}$~(see \cite{Davies}), as can be inferred from Eq.~(\ref{eq2PI.5.a}) after carrying out the $k$ integration. In this section, however, we are going to expand in powers of $m^2$, that is, the opposite to what we did in appendix~\ref{Propagators.Expansion}. However, if the result of regularisation is to be interpreted physically, it should not depend on the method used to calculate it. Indeed, that is the case: the main result of this section is that the divergences in the propagator appear $\propto m^0,\,m^2\,, m^4$, as the first terms in the $m^{2k}$ expansion, or, in the scheme of~\cite{Davies} as the only terms in $\frac{1}{m^{2k}}$ with $k<0$.

We shall consider the Dirac equation for the fermionic mode functions in cosmological space-times~\cite{Prokopec3}: 
\begin{subequations}
\begin{eqnarray}
\label{Beq1.1.1}  \frac{\text{d}U_{+,h}(\eta, k)}{\text{d}\eta} -i h k U_-(\eta, k) + i a m U_+(\eta, k) &=& 0\,,\\
\label{Beq1.1.2} \frac{\text{d}U_{-,h}(\eta, k)}{\text{d}\eta} -i h k U_+(\eta, k) - i a m U_-(\eta, k) &=& 0\,.
\end{eqnarray}
\end{subequations}
Here $U_{\pm,h}$ are the particle/antiparticle mode functions of definite helicity, given by 
\[U_{\pm,h} = R_h \pm L_h \,,\]
where $R_h, L_h$ are right and left chiralities modes of a given helicity. 

Given that $a(\eta) = \left (\frac{H_0 \eta}{2}\right )^2$ in matter era\footnote{Here the label $0$ denotes the initial value of $H(\eta)$.} and defining 
\begin{subequations}
\begin{eqnarray}
\tilde{m} &=& \frac{m H^2_0 }{4  k^3}, \\
z &=& \frac{ 2 h k }{\mathcal{H}} = k \eta,
\end{eqnarray}
\end{subequations}
we immediately find
\begin{subequations}
\begin{eqnarray}
\label{Beq1.2.1}  \frac{\text{d}U_+(z)}{\text{d}z} -i h U_-(z) + i \tilde{m} z^2 U_+(z) &=& 0\,,\\
\label{Beq1.2.2} \frac{\text{d}U_-(z)}{\text{d}z} -i h U_+(z) - i  \tilde{m} z^2 U_-(z) &=& 0\,.
\end{eqnarray}
\end{subequations}
In the case we are analysing, $\tilde{m}$ is a small parameter, because $H_0$ is suppressed by the Planck mass as one can see from Friedmann equations. In this case we can expand in powers of $\tilde{m}^k$, 
\begin{equation}
\begin{split}
\label{Beq1.3} U_+(z) =& f_0(z) + i \tilde{m} f_1(z) + \tilde{m}^2 f_2(z) + \tilde{m}^3 f_3(z)+ \tilde{m}^4 f_4(z)  + \mathcal{O}(\tilde{m}^5)\,, \\
U_-(z) =& g_0(z) + i \tilde{m} g_1(z) + \tilde{m}^2 g_2(z)+ \tilde{m}^3 g_3(z) + \tilde{m}^4 f_4(z) + \mathcal{O}(\tilde{m}^5)\,. 
\end{split}
\end{equation}
This leads to the coupled equations 
\begin{subequations}
\begin{eqnarray}
f_0'' - i g_0 &=&0 \,,\\
g_0'' - i f_0 &=&0 \,,\\
f_n'' - i g_n - (-1)^n z^2 f_{n-1} &=&0 \,,\\
g_n'' - i f_n + (-1)^n z^2 g_{n-1} &=&0 \,.
\end{eqnarray}
\end{subequations}
Which lead to the solution 
\begin{equation}
\begin{split}
\label{Beq1.6}
U_{\pm,h}(z)=&\mathcal{N} \bigg (e^{i z } \mp \frac{ \tilde{m}}{4}e^{i z } \bigg[2 z ^2 +2 i  z  - 1 \bigg ] +\\
&+\frac{i\tilde{m}^2}{120}e^{iz }  z ^3 \bigg [12  z ^2 + 15i z  - 10 \bigg ] +\mathcal{O}(\tilde{m}^3)\bigg )\,,
\end{split}
\end{equation}
where we wrote only the first terms of the expansion, since the expressions for the rest are rather cumbersome. The normalisation condition we still have to solve reads
\begin{equation}
\begin{split}
\label{Beq1.8}
&|U_+(z)|^2 + |U_-(z)|^2 = 1 = \mathcal{N}^2 \frac{128 + 8\tilde{m}^2-11 \tilde{m}^4}{64} \\
\implies &  \mathcal{N} \simeq\frac{1}{\sqrt{2}} \left (1-\frac{m^2H_0^4}{256 k^6}\right )
\end{split}
\end{equation}
Now we can expand in powers of $\frac{1}{k}$. Up to order $\frac{1}{k^5}$ the mode functions are
\begin{equation}
\begin{split}
\label{Beq1.9a}
&U_{\pm,h}(\eta)= \frac{e^{i  k \eta}}{\sqrt{2}}  \bigg\{1+\frac{iH_0^4 m^2 \eta^5\mp20  H_0^2 m \eta^2}{160 k}-\frac{ H_0^8 m^4 \eta^{10}\pm 40 i H_0^6 m^3 t^7+400 H_0^4 m^2 \eta^4\pm6400 iH_0^2 m \eta}{51200 k^2} \\
&+\frac{ -19 i H_0^8 m^4 \eta^9\pm684 H_0^6 m^3 \eta^6-960 i H_0^4 m^2 \eta^3\pm11520H_0^2 m}{184320 k^3}+\frac{181 H_0^8 m^4 \eta^8\pm5792 iH_0^6 m^3 \eta^5}{491520 k^4}\\
&+\frac{181i H_0^8m^4 \eta^7\mp 5068 H_0^6 m^3 \eta^4}{172032 k^5}+\mathcal{O}(H_0^{10}m^5)+\mathcal{O}\left(\frac{1}{k}\right)^5\bigg\}\,,
\end{split}
\end{equation}
Calculating the same divergent contribution as in appendix~\ref{Propagators.Expansion}, we find 
\begin{equation}
\begin{split}
\label{Beq1.10} &\lim_{x\rightarrow x'} \text{Tr} \left (i\tilde{S}_0^{\pm\pm}(x;x')\gamma^5\gamma^\sigma i\tilde{S}_0^{\pm\pm}(x';x)\gamma^5\gamma_\sigma\right ) = \\
=&\frac{1}{a^{D-1}} \int\frac{\text{d}^{D-1}k}{(2\pi)^{D-1}}\frac{\text{d}^{D-1}k'}{(2\pi)^{D-1}}\sum\limits_{h,h'} \bigg[ \frac{1}{2} \bigg (1- U^*_{+,h}(\eta,k)U_{-,h'}(\eta,k')-U^*_{-,h}(\eta,k)U_{+,h'}(\eta,k') \bigg )  \\
&+  \bigg (U^*_{-,h}(\eta,k)U_{-,h'}(\eta,k')- U^*_{+,h}(\eta,k)U_{+,h'}(\eta,k')\bigg ) \\
&+ i\bigg (U^*_{+,h}(\eta,k)U_{-,h'}(\eta,k') - U^*_{-,h}(\eta,k)U_{+,h'}(\eta,k')\bigg ) \bigg] \\
=& \text{Tr} \left (\tilde{S}^{\rm (reg),++}(x;x)\gamma^5\gamma^\sigma \tilde{S}^{\rm (reg),++}(x;x)\gamma^5\gamma_\sigma\right ) - \text{Tr} \left (\tilde{S}^{\rm (reg),++}(x;x)\right )^2 \\
&+\frac{1}{16 \pi^4} \left [\left ( \frac{mR}{24} + \frac{m^3}{2} \right ) \left ( \frac{2\mu^{D-4}}{D-4} + \log\left (\frac{m^2}{4\pi\mu^2}\right )\right ) +4\pi^2 \text{Tr} \left (\tilde{S}^{\rm (reg),++}(x;x)\right ) \right ]^2\,,
\end{split}
\end{equation}
which yields precisely to the counter-terms~(\ref{Counter terms}--\ref{square completion relations}),
which will eventually lead to the renormalized action~(\ref{Action and counterterms: end}).
This is a good example of the universality of quantum divergencies and their renormalization: 
the first coefficients of the expansion of the two-points function in powers of $m^{2k}$,
lead to the divergent contributions and are renormalized by the counter-terms~(\ref{Counter terms}--\ref{square completion relations}),
no matter the scheme one uses to calculate them. 
\newpage
\bibliography{bibliography}
\bibliographystyle{plain}

\end{document}